\documentclass[prb,aps,showpacs,floatfix,floats,nobalancelastpage,twocolumn]{revtex4-1}
\usepackage{graphicx}
\usepackage{graphics}
\usepackage{latexsym,amsmath,amssymb,bm,euscript}
\usepackage{color}
\usepackage{dcolumn}
\usepackage{bm}
\usepackage{epstopdf}
\usepackage{times}
\usepackage{multirow}
\usepackage{wasysym}
\usepackage{subfigure}
\def\etal{~\textit{et~al.}}
\def\ra{\rangle}
\def\la{\langle}

\def\Hc{{\rm H.c.}}
\def\fI{f_{\rm I}}
\def\fII{f_{\rm II}}

\def\ET{{$\kappa$-(ET)$_2$Cu$_2$(CN)$_3$}}
\def\dmit{{EtMe$_3$Sb[Pd(dmit)$_2$]$_2$}}

\begin{document}

\title{Majorana Spin Liquids on a two-leg ladder}
\author{Hsin-Hua Lai}
\affiliation{Department of Physics, California Institute of Technology, Pasadena, California 91125, USA}
\author{Olexei I. Motrunich}
\affiliation{Department of Physics, California Institute of Technology, Pasadena, California 91125, USA}
\date{\today}

\begin{abstract}
We realize a gapless Majorana Orbital Liquid (MOL) using orbital degrees of freedom and also an SU(2)-invariant Majorana Spin Liquid (MSL) using both spin and orbital degrees of freedom in Kitaev-type models on a 2-leg ladder. The models are exactly solvable by Kitaev's parton approach, and we obtain long-wavelength descriptions for both Majorana liquids. The MOL has one gapless mode and power law correlations in energy at incommensuare wavevectors, while the SU(2) MSL has three gapless modes and power law correlations in spin, spin-nematic, and local energy observables. We study the stability of such states to perturbations away from the exactly solvable points.  We find that both MOL and MSL can be stable against allowed short-range parton interactions.  We also argue that both states persist upon allowing $Z_2$ gauge field fluctuations, in that the number of gapless modes is retained, although with an expanded set of contributions to observables compared to the free parton mean field.
\end{abstract}
\maketitle

\section{Introduction}
Since the experimental realization of gapless quantum spin liquids (QSL)\cite{LeeNagaosaWen, Polchinski94, Altshuler94, YBKim94, SSLee2009, Metlitski2010, Mross2010, DBL, Rantner2002, Hermele2004, SenthilFisher_Z2, PWA_science, BZA, Ioffe1989, Lee08_science, Balents_nature} in two-dimensional (2D) organic compounds \ET\ and \dmit,\cite{Shimizu03, Kurosaki05, SYamashita08, MYamashita09, McKenzie, Itou07, Itou08, Itou10, MYamashita10, Powell10, SYamashita11, Kanoda11_report} there have been many theoretical proposals\cite{ringxch, LeeandLee05, Grover09, Qi09, Xu09, Biswas11} for such intriguing phases.  Among them, the proposal of an SU(2)-invariant Majorana spin liquid (MSL) by Biswas\etal\cite{Biswas11} is fascinating and in need of more careful consideration.

In an earlier work, we constructed an exactly solvable microscopic model\cite{Lai_MSL11} in Kitaev's spirit\cite{Kitaev06} to study the properties of such SU(2)-invariant MSL with Fermi surfaces of partons.  However, we allowed very low symmetries---lack of parity, inversion, and time reversal symmetry (TRS)---to sidestep discussing possible perturbations such as Cooper pairing instability, which can destabilize the gapless QSL phases away from the exactly solvable limit.  In order to study the stability of such new class of gapless QSL and further explore their properties, we realize such states on a 2-leg square ladder and show that they represent new quasi-one-dimensional (1D) phases.

We first consider a gapless Majorana Orbital Liquid (MOL) realized in a Kitaev-type model on the 2-leg ladder using orbital degrees of freedom.   The system can be reduced to one species of Majorana fermions coupled to background $Z_2$ gauge fields such that it is exactly solvable and has gapless partons with incommensurate Fermi wave vectors.  We formulate a long-wavelength description in terms of right-moving and left-moving complex fermions $f_{R/L}$ and show that local energy observable has power law correlations at incommensurate ``$2k_F$'' wavevectors.  Going away from the exactly solvable point, we first consider allowed residual parton interactions and find that there is only one valid four-fermion term and it is strictly marginal; hence, the MOL is stable to such perturbations.  

An important question is the stability of the MOL to allowing $Z_2$ gauge field fluctuations, as these lead to confinement of partons in gapped phases in so-called even $Z_2$ gauge theories in (1+1)D.\cite{SenthilFisher_Z2, Moessner01}  We argue that because of the nontrivial momenta caried by the gapless partons, there is a destructive interference for $Z_2$ vortices (instantons) in space-time, and hence these are suppressed and do not affect the count of gapless modes.  The local energy observables obtain new contributions beyond the mean field, and in this sense the partons become ``less free'', but their bosonized fields still remain very convenient for characterizing the MOL phase.

We next realize an SU(2)-invariant Majorana Spin Liquid (MSL) using both spin-1/2 and orbital degrees of freedom\cite{Fawang09,Yao10} at each site of the 2-leg ladder.\cite{Feng07}  The system can be reduced to three species of Majorana fermions coupled to background $Z_2$ gauge fields such that it is exactly solvable and has gapless partons with incommensurate wave vectors.  We formulate long-wavelength description in terms of three right-moving and left-moving complex fermions ($f^x_{R/L},~f^y_{R/L},~f^z_{R/L}$) that transform as a vector under spin rotation.  Because there is no global U(1) symmetry, in addition to familiar four-fermion residual interactions expressed as $f^{\alpha\dagger}_R f^{\beta\dagger}_L f^\gamma_R f^\delta_L$, there are other allowed terms such as $f^{\alpha\dagger}_R f^{\beta\dagger}_L f^{\gamma\dagger}_R f^{\delta\dagger}_L$. Despite of having more allowed interactions, a weak coupling renormalization group (RG) analysis gives a large regime of a stable phase. Similarly to the MOL case, we argue that such MSL with gapless matter can be also stable against $Z_2$ gauge field fluctuations even in (1+1)D.\cite{Fradkin79, Senthil99, Moessner01}

The paper is organized as follows.  In Sec.~\ref{Sec:spinless_case}, we realize the MOL with one fermion species in a Kitaev-type model\cite{Kitaev06} on the 2-leg ladder and consider its long-wavelength properties and stability against perturbations. In Sec.~\ref{Sec:SU(2)_case}, we realize the SU(2) MSL and use weak coupling RG analysis to study the stability of such phase against residual parton interactions and also discuss the stability against gauge field fluctuations.
We conclude in Sec.~\ref{sec:concl} with some discussions. In Appendix~\ref{Appendix:deconfinement}, we consider more abstractly the stability of gapless U(1) matter against $Z_2$ gauge field fluctuations in (1+1)D.
In Appendix~\ref{Appendix:SU(2)observables}, we give long-wavelength description of the SU(2) MSL and discuss observable properties. In Appendix~\ref{Appendix:Zeeman}, we consider Zeeman magnetic fields on the SU(2) MSL. 
In Appendix~\ref{Appendix:TRB}, we realize the SU(2) MSL in a model with explicitly broken time reversal symmetry and show that this case has a larger window of stability to weak perturbations.

\section{Gapless Majorana orbital liquid (MOL) on a two-leg ladder}\label{Sec:spinless_case}

\begin{figure}[t]
\subfigure[]{\label{2legladder_TRI}\includegraphics[scale=0.5]{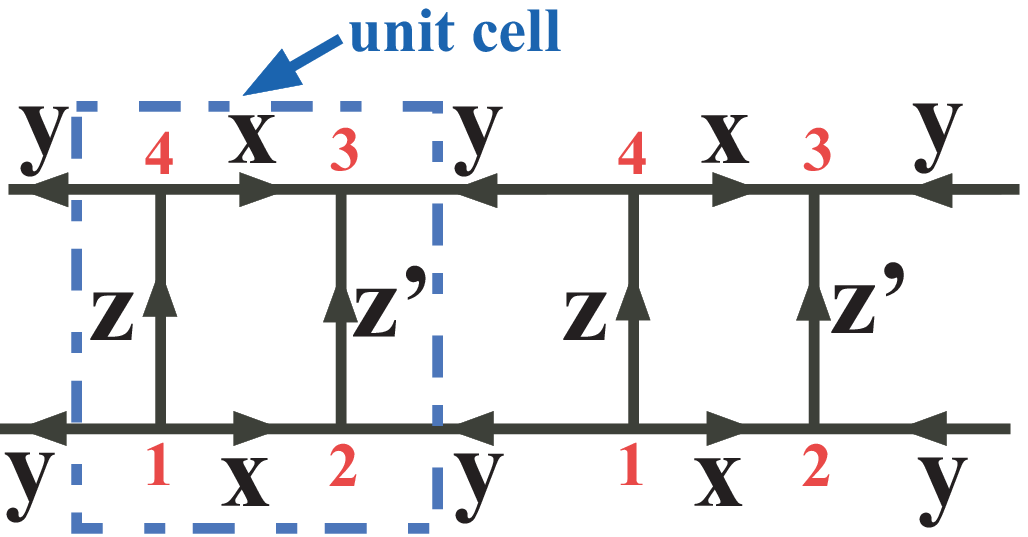}}
\subfigure[]{\label{2legladder_spec}\includegraphics[scale=0.4]{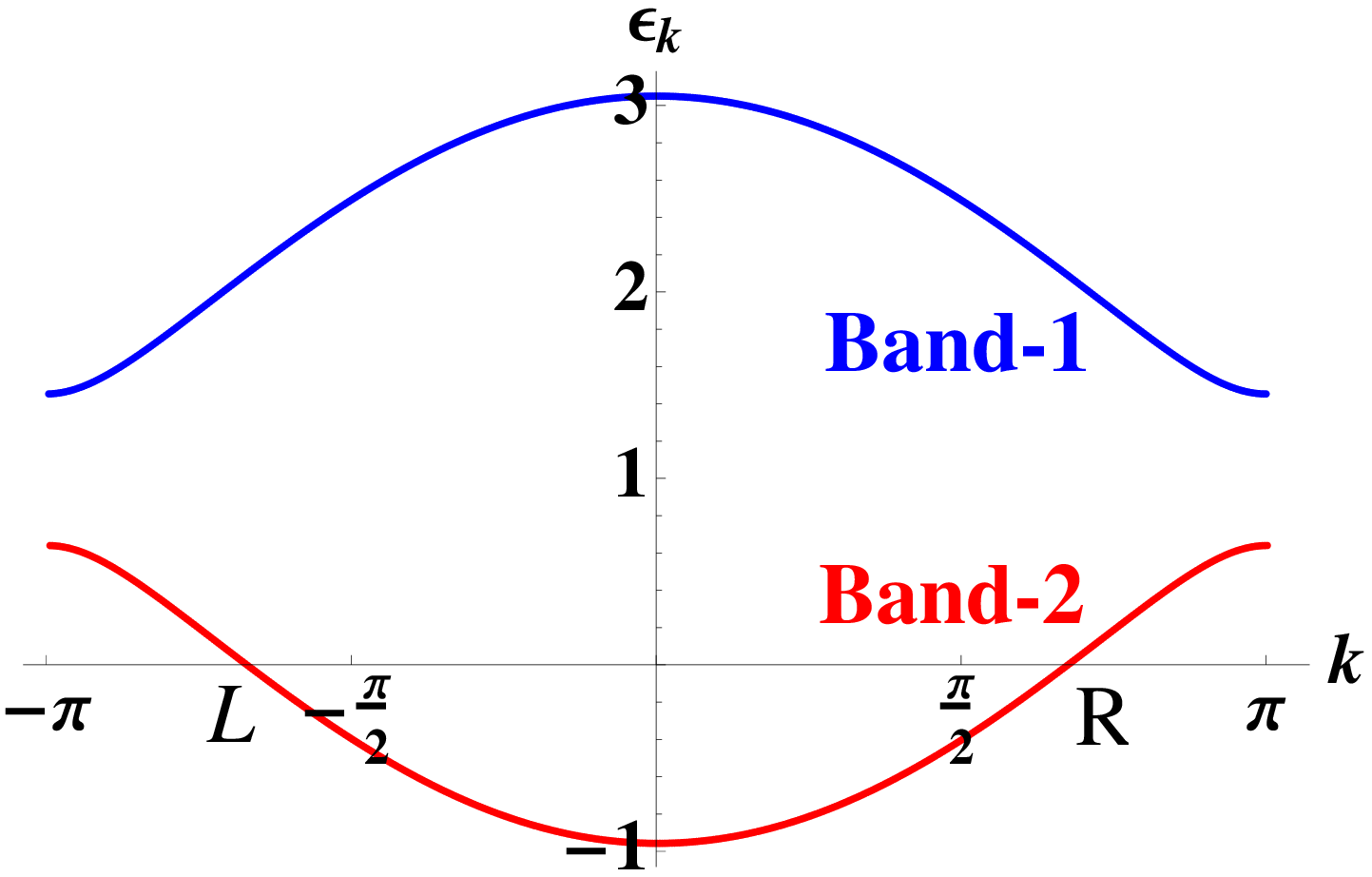}}
\caption{(a) Graphical representation of the exactly solvable Kitaev-type model on the 2-leg ladder and its solution in the zero flux sector.  The $c$ Majoranas propagate with pure imaginary hopping amplitudes specified by the couplings $J_x,~J_y,~J_z,$ and $J'_z $; the signs in our chosen gauge are indicated by the arrows and the four-site unit cell is also indicated. (b) Dispersion of complex fermions that solve the Majorana problem for parameters $ \{J_x,~J_y,~J_z ,~J'_z \}= \{1.2,~0.8,~1.0,~1.1 \} $.}
\label{TRInvariant}
\end{figure}

We begin with a ``spinless'' (one species) MOL realized in a Kitaev-type model on a 2-leg ladder shown in Fig.~\ref{2legladder_TRI}. 
The Hamiltonian is 
\begin{eqnarray}
\mathcal{H}=\mathcal{H}_0+K_{\square_{xz}}\sum_{\square_{xz}}W_{\square_{xz}}+K_{\square_{yz}}\sum_{\square_{yz}}W_{\square_{yz}} ~,
\label{HMOL}
\end{eqnarray}
where
\begin{eqnarray}
&& \mathcal{H}_0=\sum_{\lambda-{\rm link}, \la j k \ra}J_{jk}\tau^{\lambda}_{j}\tau^{\lambda}_{k},~ \label{H0}\\
&& W_{\square_{xz}}=\tau^{y}_1\tau^{y}_2\tau^{y}_3\tau^{y}_4,~\label{Wterm_xz}\\
&& W_{\square_{yz}}=\tau^x_2\tau^x_1\tau^x_4\tau^x_3.~\label{Wterm_yz}
\end{eqnarray}
The $\vec{\tau}$ Pauli matrices can be thought of as acting on two-level orbital states.  The $W_p$ terms, with $p=\square_{xz}$ or $\square_{yz}$ formed by $x$ and $z$ or $y$ and $z$ links respectively, are plaquette operators which commute among themselves and with all other terms in the Hamiltonian and are added to stabilize particular flux sector, see Fig.~\ref{2legladder_TRI}.  Following Kitaev's approach, we introduce Majorana representation as
\begin{eqnarray}\label{orbital majorana reps}
\tau^{\alpha}_j=ib^{\alpha}_j c_j,~
\end{eqnarray}
with the constraint $D_j \equiv b^x_j b^y_j b^z_j c_j = 1$. The Hamiltonian can be rephrased as
\begin{eqnarray}
&& \mathcal{H}_0 = i \sum_{\la j k \ra} \hat{u}_{jk} J_{jk} c_j c_k ~,\\
&& W_{p=\{\square_{xz}, \square_{yz}\}}=-\prod_{\la jk \ra\in p} \hat{u}_{jk} ~,\label{W-terms}
\end{eqnarray}
where $\hat{u}_{jk} \equiv -ib^{\lambda}_j b^{\lambda}_k$ for $\lambda$-link $\la j k \ra$ and the product in the last line is circling the plaquette.

Following familiar analysis in Kitaev-type models, we observe that in the enlarged Hilbert space, $\hat{u}_{jk}$ commute among themselves and with the Hamiltonian, and we can proceed by replacing them by their eigenvalues $\pm 1$ and interpreting as static $Z_2$ gauge fields.  The $W_{p}$ terms, with $K_p > 0$ assumed to be sufficiently large, can be used to stabilize the sector with zero fluxes through all elementary plackets, and this can give a gapless phase.  In our work, we fix the gauge by taking $u_{jk}=1$ for bonds $j\rightarrow k$ as shown by the arrows in Fig.~\ref{2legladder_TRI}. 

There are four physical sites per unit cell, so there are four Majoranas per unit cell.  From now on, we replace the site labeling $j$ with $j=\{ X,a\}$, where $X$ runs over the one-dimensional lattice of unit cells of the ladder and $a$ runs over the four sites in the unit cell, see Fig.~\ref{2legladder_TRI}.  The Hamiltonian can be written as,
\begin{eqnarray*}
\mathcal{H} = \sum_{\la jk \ra} c_{j} \mathcal{A}_{jk} c_{k}
= \sum_{\la (X,a), (X',a')\ra} c_{X,a} \mathcal{A}_{X,a; X',a'} c_{X',a'}.\end{eqnarray*}
There is translational symmetry between different unit cells, and $\mathcal{A}_{ X,a; X',a'} = \mathcal{A}_{aa'}(X - X')$.

In order to give a concise long-wavelength description, it will be convenient to use familiar complex fermion fields.  To this end, we can proceed as follows.  For a general Majorana problem specified by an anti-symmetric pure imaginary matrix $\mathcal{A}_{jk}$, we diagonalize $\mathcal{A}_{jk}$ for spectra, but only half of the bands are needed while the rest of the bands can be obtained by a specific relation and are redundant.  Explicitly, for a system with $2m$ bands, we can divide them into two groups.  The first group contains bands from $1$ to $m$ with eigenvector-eigenenergy pairs $\{\vec{v}_{b, k}, \epsilon_{b, k}\}$, where $b = 1, 2, \dots, m$ are band indices, and the second group contains bands from $m+1$ to $2m$ related to the first group, $\{\vec{v}_{b'=m+b,  k}, \epsilon_{b'=m+b, k}\} = \{ \vec{v}^*_{b, -k}, -\epsilon_{b, -k}\}$. Using only the bands with $b=1$ to $m$, we can write the original Majoranas in terms of usual complex fermions as
\begin{eqnarray*}\label{usual_fermion}
c(X,a) = \sqrt{\frac{2}{N_{uc}}}\sum_{b=1}^m \sum_{k \in {\bf B.Z.}}\left[ e^{i k X} v_{b,k}(a) f_{b}(k) +\Hc \right],
\end{eqnarray*}
where $N_{uc}$ is the number of unit cells, ${\bf B.Z.}$ stands for the Brillouin Zone, and the complex fermion field $f$ satisfies the usual anticommutation relation, $\{ f^\dagger_{b}(k),f_{b'}(k')\}=\delta_{b b'}\delta_{kk'}$. In terms of the complex fermion fields, the Hamiltonian becomes
\begin{eqnarray}\label{usual_fermion_H}
\mathcal{H}=\sum_{b=1}^{m}\sum_{k\in {\bf B.Z.}} 2\epsilon_{b}(k) \left[ f^{\dagger}_{b}(k)f_b (k)-\frac{1}{2}\right] .
\end{eqnarray}
In the present case, $2m = 4$ and therefore two bands are sufficient to give us the full solution of the Majorana problem. 

The above approach can be applied to any general Majorana problem and is needed when we consider a model lacking any symmetries in Appendix~\ref{Appendix:TRB}.  In the present case, we require the model to respect time reversal symmetry\cite{Kitaev06} and leg interchange symmetry, which allows us to introduce convenient complex fermion fields already on the lattice scale as follows
\begin{eqnarray}
\fI(X) &=& \frac{c(X,1) + i c(X,4)}{2} ~, \label{fI}\\
\fII(X) &=& \frac{-i c(X,2) + c(X,3)}{2} ~. \label{fII}
\end{eqnarray}
The Hamiltonian becomes
\begin{eqnarray*}
\mathcal{H} &=& 2\sum_X \bigg{\{} J_z \fI^\dagger(X) \fI(X) + J^\prime_z \fII^\dagger(X) \fII(X) - \\
&& -\left[ J_x \fI^\dagger(X) \fII(X) + J_y \fII^\dagger(X) \fI(X+1) + \Hc \right] \bigg{\}},
\end{eqnarray*}
where we ignored constant contribution.  It is easy to calculate the band dispersions,
\begin{equation}
\epsilon(k) = J^{+}_z \pm \sqrt{(J^{-}_z)^2 + J_x^2 + J_y^2 + 2 J_x J_y \cos(k)},
\end{equation}
with $J^{\pm}_z = (J_z \pm J^\prime_z)/2$. The spectrum is gapless for $|J_x - J_y| \leq \sqrt{J_z J_z^\prime} \leq J_x + J_y$, where without loss of generality we assumed all couplings to be positive. For an illustration of the energy spectrum, we take $\{ J_x,~J_y,~J_z,~J^\prime_z\} = \{1.2,~0.8,~1.0,~1.1\}$ and show the two bands of the complex fermions in Fig.~\ref{2legladder_spec} labeled from top to bottom as band-1 and band-2.  We note that the gapless phase occur in a large parameter regime and there is no fine tuning here.  The specific parameters are chosen to emphasize that we do not require any symmetries other than time reversal and leg interchange.

The band-2 crosses zero at $k_{FR}$ and $k_{FL} = -k_{FR}$ from time reversal. For long wavelength physics, we can focus on this band and introduce continuum complex fermion fields $f_{R/L}$; for the lattice Majoranas, we obtain the expansion,
\begin{eqnarray}\label{continuum_expansion}
c(X, a) \sim \sum_{P=R/L} \left[ e^{i k_{FP} X} v_{2,P}(a) f_P(X) + \Hc \right] ~.
\end{eqnarray}
From the detailed band calculation, at the right Fermi point 
\begin{eqnarray}
\vec{v}_{2,R} = \sqrt{\frac{J^\prime_z}{4J^+_z}}
\begin{pmatrix}
1 \\
i \xi \\
\xi\\ 
-i
\end{pmatrix} ~,
\end{eqnarray}
where $\xi = (J_x + J_y e^{i k_{FR}})/J^\prime_z$.  Using time reversal invariance, for the left Fermi point we get $v_{2,L}(a) = (-1)^{a+1} v^{*}_{2,R}(a)$.  The effective low energy Hamiltonian density is
\begin{eqnarray}
\mathcal{H} = v_F \left[f^\dagger_R (-i \partial_x) f_R - f^{\dagger}_L  (-i \partial_x) f_L\right] ~,
\end{eqnarray}
describing a one-dimensional Dirac particle with Fermi velocity $v_F = J_x J_y \sin(k_{FR})/J^{+}_z$.  We list the symmetry transformations of the continuum fields in Table~\ref{tab:transfprops} (ignoring the ``spin'' indices there).  
In particular, the leg interchange symmetry prohibits terms of the form $f_R f_L$ from the continnum Hamiltonian that would gap out the spectrum.

\subsection{Fixed-point theory of Majorana orbital liquid and observables}\label{MOL:observable}
In this subsection, we first give the fixed-point theory of the MOL and then we will consider bond energy operators to characterize such gapless phase. We use Bosonization, re-expressing the low-energy fermion operators with Bosonic fields,\cite{Shankar_Acta, Lin97, Marston02}
\begin{eqnarray}
f_P = e^{i (\varphi + P \theta)} ~,
\end{eqnarray}
with canonical conjugate boson fields:
\begin{eqnarray}
&& \left[\varphi(x), \varphi(x')\right]=\left[\theta(x),\theta(x')\right]=0 ~,\label{spinless-comut1}\\
&& \left[\varphi(x), \theta(x')\right]=i\pi \Theta(x-x') ~,\label{spinless-comut2}
\end{eqnarray}
where $\Theta(x)$ is the Heaviside step function.

The fixed-point bosonized Lagrangian of such gapless MOL is 
\begin{eqnarray}\label{fixed-point-theory:MOL}
\mathcal{L}_{MOL}=\frac{1}{2\pi g} \left[\frac{1}{v} (\partial_\tau \theta)^2+v (\partial_x \theta)^2\right].~
\end{eqnarray}
For free fermions, $g=1$ and $v=v_F$, the bare Fermi velocity.  Later when we discuss the stability of such a phase in Sec.~\ref{MOL:stability}, we will see that there is only one strictly marginal interaction which introduces one Luttinger parameter $g$.  To detect the gaplessness of the phase using physical (gauge-invariant) observables, here we consider bond-energy operators,\cite{Lai11_bondcorr} $\mathcal{B}^{s/a}(X)$, which we further categorize into symmetric or anti-symmetric with respect to the leg interchange symmetry. The specific microscopic operators are
\begin{eqnarray}
\nonumber 
&&\mathcal{B}^{s/a}(X) = \tau^x(X,1)\tau^x(X,2)\pm\tau^x(X,4)\tau^x(X,3)\\
&&\hspace{1cm}= i u_{12}c(X,1)c(X,2) \pm iu_{43}c(X,4)c(X,3), ~~~~~~ \label{Bond_energy}
\end{eqnarray}
where we used Majorana representation, Eq.~(\ref{orbital majorana reps}). In our gauge, after expansion in terms of the continuum complex fermions using Eq.~(\ref{continuum_expansion}), the Fourier components are organized as follows
\begin{eqnarray}
&&\mathcal{B}^{s}_{Q=0} \sim f^{\dagger}_{R}f_{R}+f^{\dagger}_{L}f_{L}=\frac{\partial_x \theta}{\pi},~\\
&& \mathcal{B}^{s}_{k_{FR}-k_{FL}} \sim f^{\dagger}_L f_R = i e^{i2\theta},~\\
&& \mathcal{B}^{a}_{k_{FR}+k_{FL}} \sim f_L f_R = -i e^{i 2 \varphi}.~
\end{eqnarray}
[Note that with TRS, the wave vector $k_{FR} + k_{FL}$ is the same as $Q=0$; to be more precise, we should write a Hermitian and time reversal symmetric combination, $\mathcal{B}^{a}_{Q=0} = i f_L f_R + \Hc$]
Thus, the symmetric bond-energy correlations are expected to decay with oscillations at incommensurate wave vectors $\pm 2k_{FR}$, while the anti-symmetric bond-energy correlations decay without oscillations.  Such a sharp difference can be confirmed in exact numerical calculations.

In the bosonized form, the scaling dimension of each term is apparent, 
\begin{eqnarray}
&& \Delta[\mathcal{B}^{s}_{Q=0}] = 1, \\
&& \Delta[\mathcal{B}^{s}_{2k_{FR}}] = g, \\
&& \Delta[\mathcal{B}^{a}_{Q=0}] = \frac{1}{g}.
\end{eqnarray}
In the non-interacting parton limit, $g \rightarrow 1$, we expect to see all components of bond-energy correlations decay as $X^{-2}$. 

\begin{figure}[t]
\subfigure[Leg-symmetric energy correlation]{\label{symmB}\includegraphics[scale=0.5]{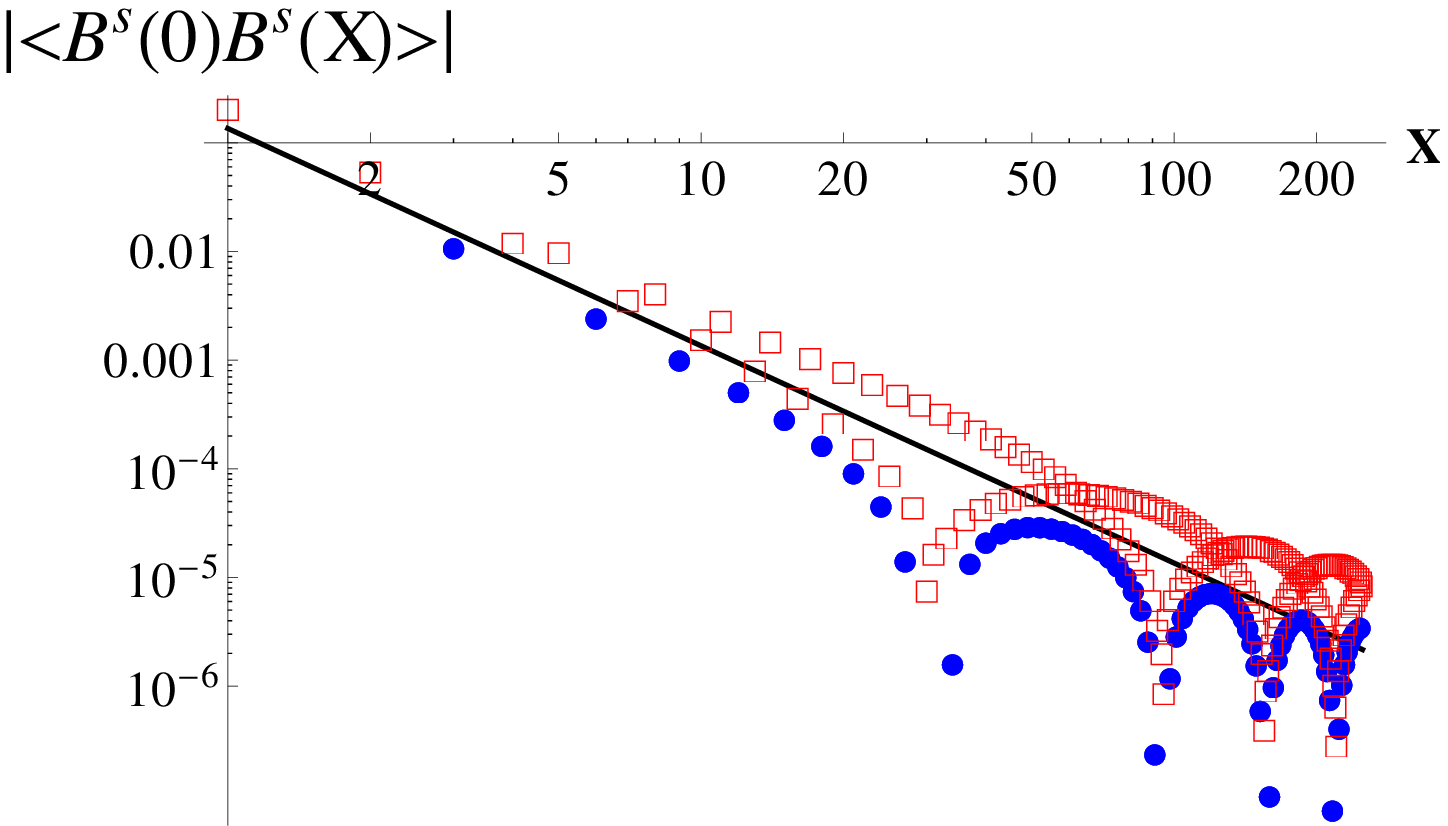}}
\subfigure[Leg-anti-symmetric energy correlation]{\label{antisymmB}\includegraphics[scale=0.5]{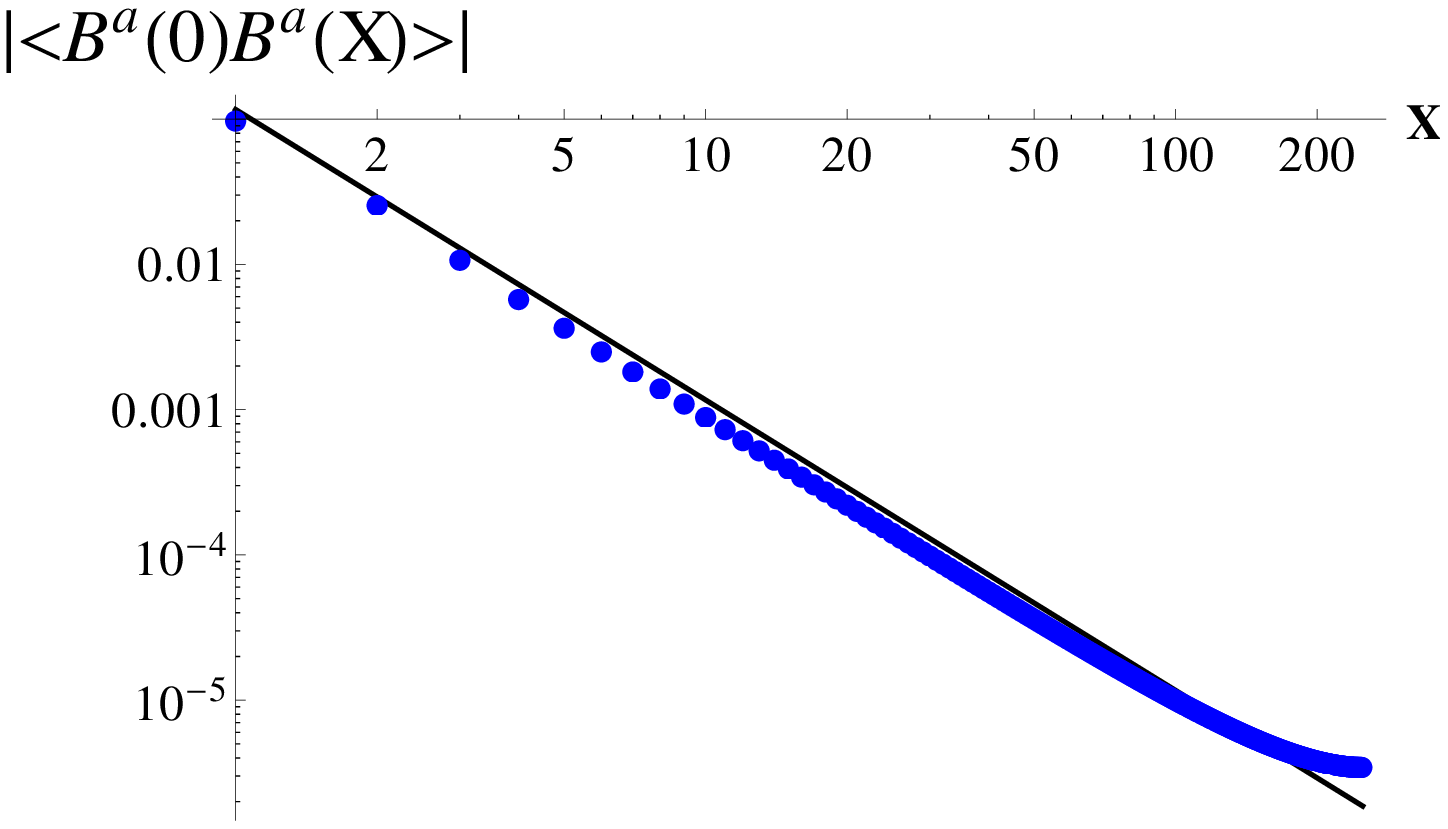}}
\caption{Figures (a) and (b) illustrate power law behaviors of the symmetric and anti-symmetric bond energy correlations, with $B^{s/a}$ defined in Eq.~(\ref{Bond_energy}), in the exactly solvable model with non-interacting partons. The system has $500$ unit cells and we use the same parameters as in Fig.~\ref{TRInvariant}. We plot absolute values and indicate the sign with filled circles (blue) for positive correlations and open square boxes (red) for negative correlations.
The log-log plots clearly show $X^{-2}$ decay (straight lines) with incommensurate oscillations in the symmetric case and no oscillations in the anti-symmetric case.  The characteristic wavevectors can be determined from the structure factor study shown in Fig.~\ref{Bond_energy_structure_factor}.
}
\label{spinless-bond}
\end{figure}

For illustration, we calculate correlations in the exactly solvable model, taking the same parameters as in Fig.~\ref{TRInvariant}. Figure~\ref{symmB} shows log-log plot of symmetric bond-energy correlations in a finite system with $500$ unit cells, while Figure~\ref{antisymmB} shows anti-symmetric bond-energy correlations.\cite{BC}  We can see the overall $X^{-2}$ envelope in both figures and also incommensurate oscillations in the symmetric bond-energy correlations, which confirm the theoretical analysis above.

\begin{figure}[t]
\subfigure[Leg-symmetric energy structure factor]{\label{structure_factor}\includegraphics[scale=0.5]{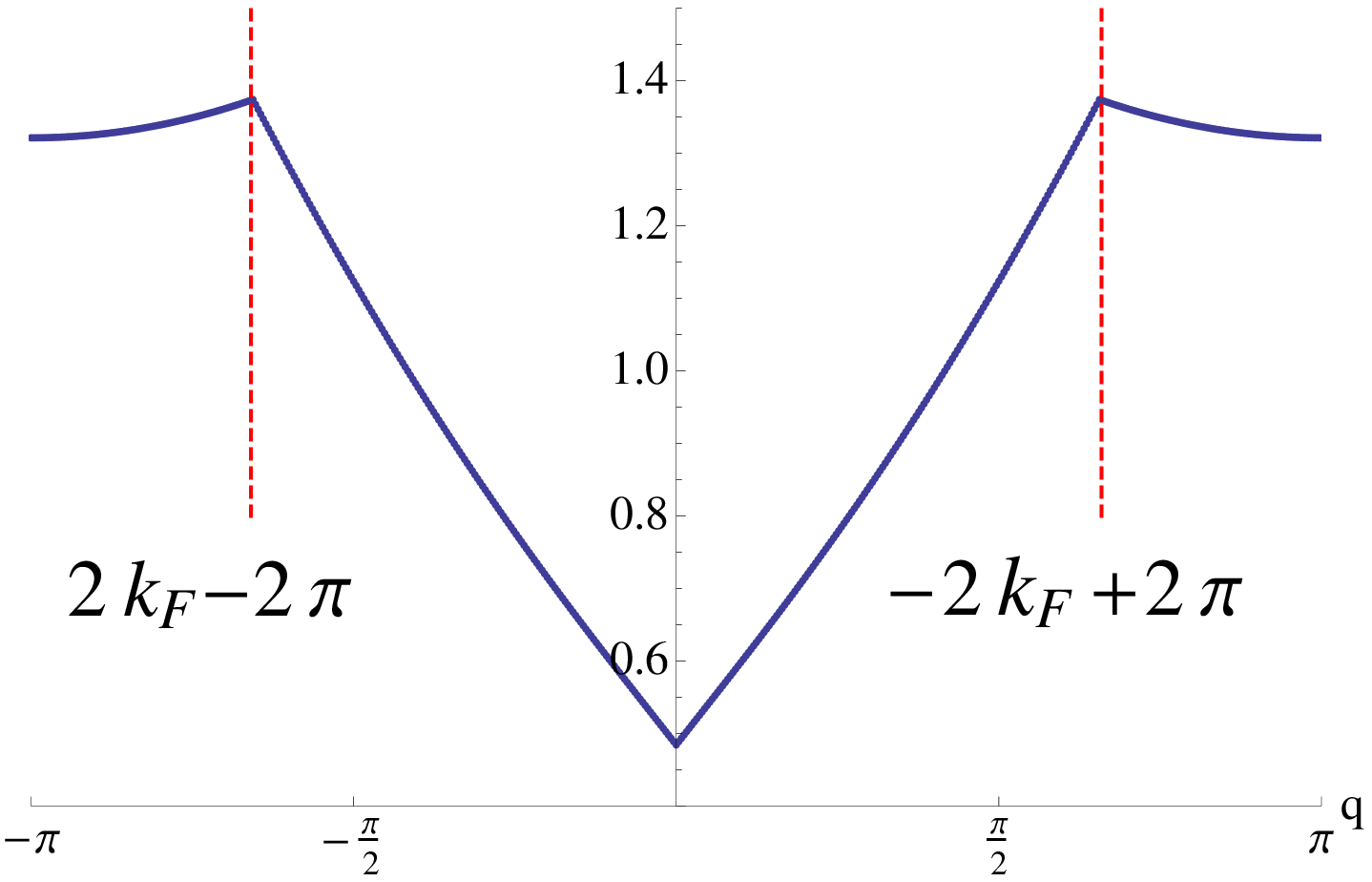}}
\subfigure[Leg-anti-symmetric energy structure factor]{\label{antisymmstructure_factor}\includegraphics[scale=0.5]{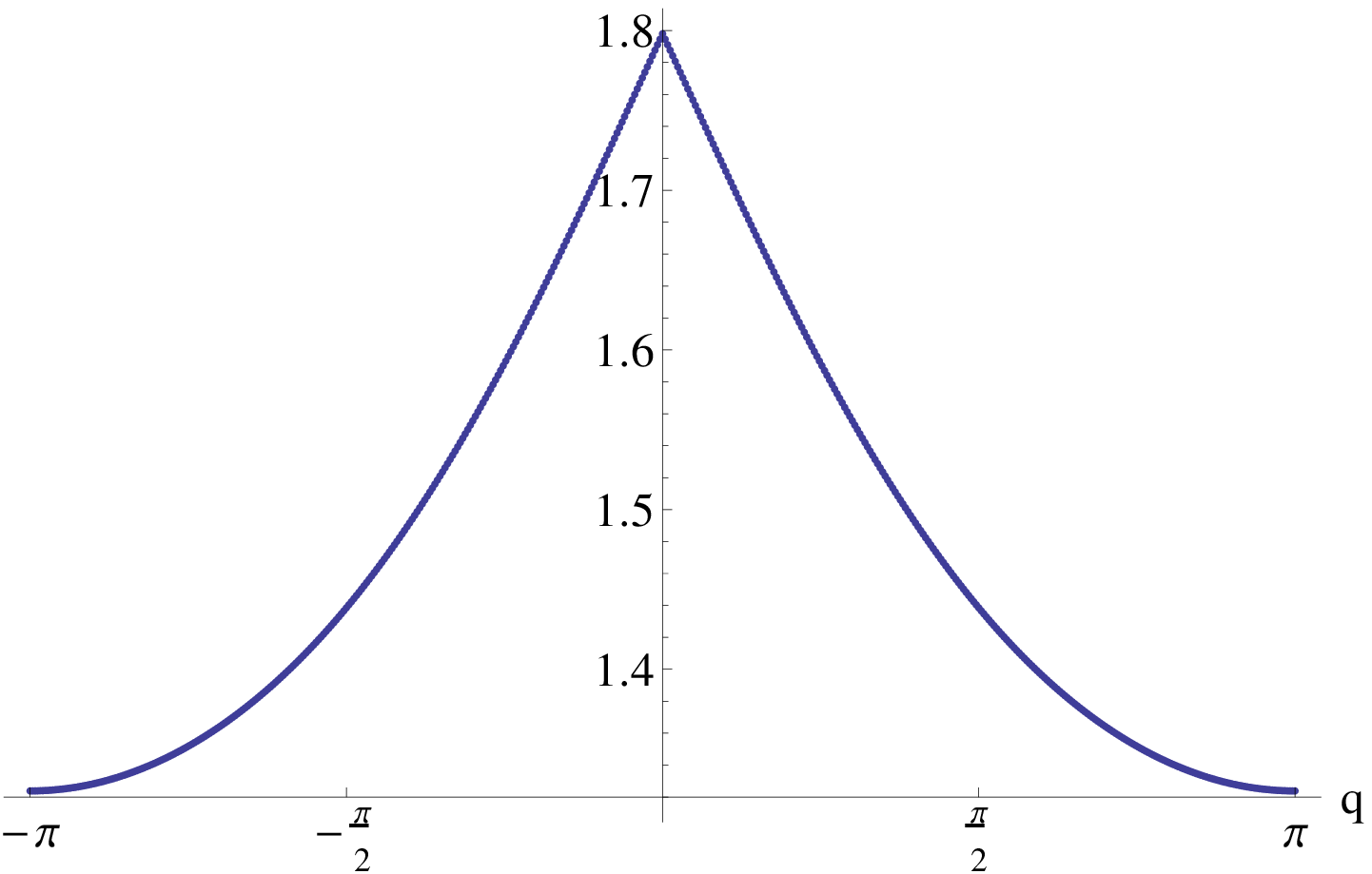}}
\caption{Figures (a) and (b) illustrate the symmetric bond-energy and anti-symmetric bond-energy structure factors corresponding to Figs.~\ref{symmB} and \ref{antisymmB} respectively.  Both cases clearly show a singularity at $Q=0$, while the symmetric case also shows singularities at $\pm 2k_{F}$.
}
\label{Bond_energy_structure_factor}
\end{figure}

Power-law correlations in real space correspond to singularities in momentum space, which we can study by considering the corresponding structure factors. Figure~\ref{structure_factor} shows the symmetric bond-energy structure factor and Fig.~\ref{antisymmstructure_factor} shows the anti-symmetric bond-energy structure factor.  It is clear that the singularities in the symmetric case occur exactly at $Q=0$ and $Q = \pm (k_{FR}-k_{FL}) = \pm 2k_{FR}\equiv \pm 2k_{F}$ (which we also mark using values obtained by extracting the Fermi points of band 2), while there is only $Q=0$ singularity for the anti-symmetric case.

Let us now consider some other operators similar to generic XYZ energy terms but not present in the exactly solvable model; this will be also useful for the subsequent discussion of the MOL stability.  First, operators like $\tau^y(X,1) \tau^y(X,2)$ and $\tau^z(X,1) \tau^z(X,2)$ have ultra-short-ranged correlations as they contain unpaired localized $b$-fermions. It is more interesting to consider operators like $\tau^x(X,1) \tau^x(X,4)$ defined on the $z$-type (vertical) links in Fig.~\ref{TRInvariant}. In this case, even though the local operator contains unpaired $b$-Majoranas, in the physical Hilbert space these can actually be paired at the expense of introducing a string product of the gapless $c$-Majoranas.  For example, consider calculating correlation between rungs at $X$ and $X'$:
\begin{eqnarray}
\nonumber && \hat{\cal F}(X,X') \equiv \tau^x(X,1) \tau^x(X,4) ~ \tau^x(X',1) \tau^x(X',4) = \\
\nonumber && = \prod_{X \leq X'' < X'} (-1) c(X'',1) c(X'',4) c(X'',2) c(X'',3) \times~~~~~ \\
&& \hspace{1cm} \times \prod_{\la (X,1),(X,4) \ra ~<~ \lambda-{\rm link}~\la ij \ra ~\leq~ \la (X',1), (X',4) \ra} \hat{u}^\lambda_{ij}
\label{rungXX_rungXX}
\end{eqnarray}
where the last product contains all links on the ladder located between the two vertical links excluding $\la (X,1), (X,4) \ra$ and including $\la (X',1), (X',4) \ra$ and oriented  as shown in Fig.~\ref{TRInvariant}.  The second line is $1$ in our chosen gauge, and we then have a factor of 
\begin{eqnarray}
\nonumber (-1) c(X'',1) c(X'',4) c(X'',2) c(X'',3) = \\
= e^{i\pi [\fI^\dagger(X'') \fI(X'') + \fII^\dagger(X'') \fII(X'')]} 
\end{eqnarray}
for each unit cell, where we used Eqs.~(\ref{fI})-(\ref{fII}). In the present gauge, we can write schematically 
$\tau^x(X,1) \tau^x(X,4) \sim \prod_{X'' < X} (-1) c(X'',1) c(X'',4) c(X'',2) c(X'',3)$, and see that this contains non-local Jordan-Wigner-like string operator in terms of the gapless partons.  In the bosonization language, the string operator becomes 
\begin{eqnarray}\label{Orbital:string}
\prod_{X'' < X} e^{i\pi [\fI^\dagger(X'') \fI(X'') + \fII^\dagger(X'') \fII(X'')]} \sim e^{\pm i[\theta(X) + \pi \bar{n} X]}.~~~~~~
\end{eqnarray}
This has scaling dimension $1/4$ in the free-fermion case and hence the above correlation decays as $X^{-1/2}$ power law and oscillates at wavevector $\pi \bar{n} = k_F$ from Fig.~\ref{2legladder_spec}.  It may seem unusual that this appears to contain the specific gauge-dependent quantity $k_F$; note, however, that in the full calculation we used the specific gauge to set the last line in Eq.~(\ref{rungXX_rungXX}) to unity, and the final result is independent of the gauge.

\begin{figure}[t] 
   \centering
   \includegraphics[scale=0.5]{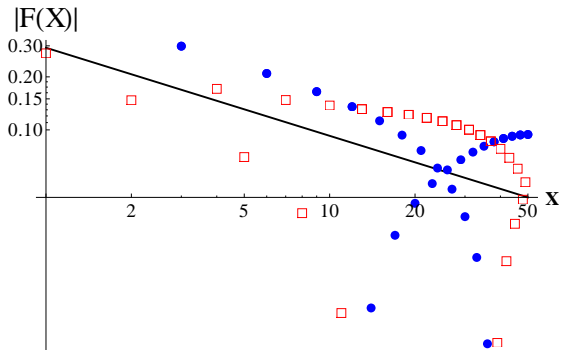} 
   \caption{Figure illustrates power-law behavior of the correlation ${\cal F}(X-X') = \la \hat{\cal F}(X,X') \ra$, defined in Eq.~(\ref{rungXX_rungXX}).  The system has $100$ unit cells in chain length and the same parameters as in Fig.~\ref{TRInvariant}.   We show the absolute values of $|{\cal F}(X)|$ and indicate the sign with filled circles (blue) for positive correlations and open square boxes (red) for negative correlations.  The log-log plot clearly shows $X^{-1/2}$ envelope (straight line in the figure).  The irregular behavior is due to incommensurate oscillations.
} 
   \label{loglogstring}
\end{figure}

\begin{figure}[t] 
   \centering
   \includegraphics[scale=0.5]{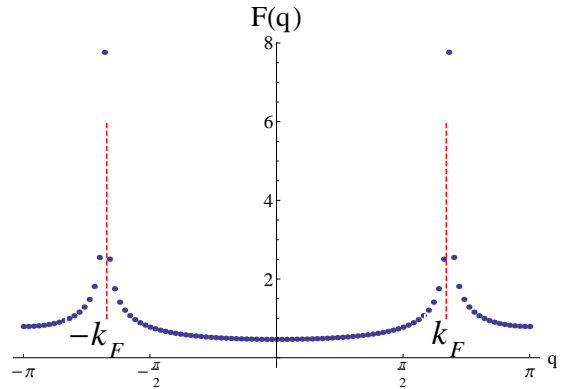} 
   \caption{Structure factor corresponding to Fig.~\ref{loglogstring}; we also mark the expected locations of the singularities, $\pm k_F$.
}
   \label{Fig:string_structure_factor}
\end{figure}

Evaluating expectation value of the string operator in the free fermion ground state leads to a Pfaffian of a matrix formed by the Majorana contractions and can be easily computed numerically for reasonable sizes.\cite{Wimmer_M11}  The results are shown in Fig.~\ref{loglogstring} for a system with $100$ unit cells.\cite{BC} The corresponding structure factor is shown in Fig.~\ref{Fig:string_structure_factor}.  We can clearly see the singularities at $\pm k_F$ and confirm our theoretical analysis.

\subsection{Stability of Majorana orbital liquid}\label{MOL:stability}
Let us now consider going away from the exactly solvable point.  First, we consider perturbations that are local in the continuum fermion fields. This ignores fluctuations in the $Z_2$ gauge fields, and we will address stability against confinement shortly. 
In the language of usual complex fermions, there is only one valid 4-fermion interaction,
\begin{eqnarray}
\mathcal{H}_{\rm int} = u f^{\dagger}_{R} f_{R} f^{\dagger}_{L} f_L ~.
\end{eqnarray}
This interaction is strictly marginal, and therefore the gapless MOL is stable also with $\mathcal{H}_{\rm int}$ and has one gapless mode.  This interaction will renormalize the Luttinger parameter and the Fermi velocity to be 
\begin{eqnarray}
&& g = \sqrt{ \frac{1 - \frac{u}{2\pi v_F}}{1 + \frac{u}{2\pi v_F}} } ~, \\
&& v = v_F \sqrt{1 - \left(\frac{u}{2\pi v_F}\right)^2} ~,
\end{eqnarray}
which completes our description of the fixed-point theory in Eq.~(\ref{fixed-point-theory:MOL}) and will modify the power laws of various correlations as discussed above in Sec.~\ref{MOL:observable}.

We now want to address the issue of confinement, more precisely, the stability of the MOL theory when we allow fluctuations in the $Z_2$ gauge fields.  As we discuss in Appendix~\ref{Appendix:deconfinement}, allowing $Z_2$ gauge field fluctuations in the (1+1)D space-time is like allowing half-vortices in the phase field in the bosonized harmonic liquid description and corresponds to allowing terms $\lambda_{1/2} \cos(\theta + k_F X + \alpha_{1/2})$ in the dual harmonic liquid description, Eq.~(\ref{fixed-point-theory:MOL}).  The key point is that this term is oscillating for generic $k_F$ and hence averages out to zero (the underlying physics is destructive interference due to Berry phases).  Thus, our gapless MOL with incommensurate momenta carried by the fermion fields persists also in the presence of $Z_2$ gauge field dynamics even in (1+1)D, in the sense that we retain the gapless mode.

One may worry about the precise connection between the present system and the schematic $Z_2$ gauge theory plus $U(1)$ matter at incommensurate density considered in Appendix~\ref{Appendix:deconfinement}.  Indeed, the connection is only crude, and we do not have one-to-one correspondences.  Nevertheless, we can bolster our argument by considering explicitly some allowed perturbations to the exactly solvable model.  
Consider, e.g., adding small general XYZ interactions $\sum_{\la ij \ra} \sum_{\mu=x,y,z} \delta J^\mu_{ij} \tau_i^\mu \tau_j^\mu$ on all bonds in a manner respecting the underlying lattice symmetries.
As we have discussed earlier, $\delta J^{y,z}$ terms on the $x$-type bonds and $\delta J^{x,z}$ terms on the $y$-type bonds have short-range correlations and hence constitute irrelevant perturbations (of course, they can renormalize the Luttinger parameter).  On the other hand, $\delta J^{x,y}$ terms on the $z$-type bonds have power law correlations.  However, these correlations oscillate at the incommensurate wavevector, see Fig.~\ref{loglogstring} and Fig.~\ref{Fig:string_structure_factor}.  Hence such terms, whose structure is similar to $\lambda_{1/2} \cos(\theta + k_F X + \alpha_{1/2})$, cf.~Eq.~(\ref{Orbital:string}), are washed out from the low energy Hamiltonian.  Thus, the fixed point description is the same as described earlier, but with the additional remark that now generic energy correlations that are symmetric under the leg interchange will also obtain a contribution oscillating at wavevector $k_F$ with scaling dimension $g/4$.

Finally, we remark that the $Z_2$ gauge fluctuations do lead to 
confinement in our 2-leg model in gapped regimes, e.g., when the $J_z$ 
terms dominate over the $J_x$, $J_y$ terms in the original Hamiltonian 
Eq.~(\ref{HMOL}).  In this regime, we can start with effective 
(super)-spins on the rungs formed by the large $J_z$ terms (e.g., after 
conveniently making the $J_z$ coupling ferromagnetic).  We 
perturbatively derive effective Hamiltonian governing these effective 
spins, which works out to be an Ising-like chain and has two degenerate 
ground states.  Adding the $\delta J^{x,y}$ perturbations on the 
$z$-type bonds gives local \emph{longitudinal} fields in this Ising 
chain and immediately lifts the degeneracy.  Hence, there is a unique 
ground state.

Furthermore, creating a single domain-wall-like excitation, which 
behaves as a free particle in the exactly solvable model, requires 
infinite energy in the presence of the longitudinal field.  On the other 
hand, a pair of domain walls, kink and anti-kink, are allowed, but to 
separate one from the other requires energy linearly proportional to the 
distance between them.  Therefore, such $\delta J^{x,y}$ perturbations 
on the $z$-type bonds give linear confinement of particles that were 
free at the exactly solvable point, and this applies to all particles 
that carry gauge charge with respect to the $Z_2$ gauge field in the 
exactly solvable model.

\section{Gapless SU(2)-invariant Majorana spin liquid (MSL) on the two-leg ladder}\label{Sec:SU(2)_case}
We now want to consider Majorana spin liquids with more degrees of freedom, in particular with physical spin degrees of freedom, and see what new issues and features arise in this case.
In order to construct spin SU(2)-invariant Kitaev-type model, we follow Refs.~\onlinecite{Fawang09, Yao10, Lai_MSL11} to take a system with both spin and orbital degrees of freedom on each site.  The complete Hamiltonian is
\begin{eqnarray}
\mathcal{H}_{SU(2)}=\mathcal{H}'_0 +K_{\square_{xz}}\sum_{\square_{xz}}W_{\square_{xz}}+K_{\square_{yz}}\sum_{\square_{yz}}W_{\square_{yz}},~
\end{eqnarray}
where
\begin{eqnarray}
&& \mathcal{H}'_0=\sum_{\lambda-{\rm link}, \la j k \ra} J_{jk} \left(\tau^{\lambda}_{j}\tau^{\lambda}_{k}\right)\left( \vec{\sigma}_{j} \cdot \vec{\sigma}_k \right) ~.
\end{eqnarray}
$\mathcal{H}'_0$ is a Kugel-Khomskii-like Hamiltonian with $\vec{\sigma}$ being the spin-1/2 Pauli matrices and $\vec{\tau}$ being the Pauli matrices acting on the orbital states, while the $W_{\square_{xz}}$ and $W_{\square_{yz}}$ terms are given in Eqs.~(\ref{Wterm_xz})-(\ref{Wterm_yz}). 
  
Introducing Majorana representation of spin-1/2, we write the spin and orbital operators as
\begin{eqnarray}
\sigma^{\alpha}_j=-\frac{i}{2}\sum_{\beta,\gamma}\epsilon ^{\alpha \beta \gamma} c^\beta_j c^\gamma_j ~, \label{Majorana_reps:spin}\\
\tau^{\alpha}_j=-\frac{i}{2}\sum_{\beta,\gamma}\epsilon ^{\alpha \beta \gamma} d^\beta_j d^\gamma_j ~. \label{Majorana_reps:orbital}
\end{eqnarray}
On each site $j$ of the 2-leg ladder, we realize the physical four-dimensional Hilbert space using six Majorana fermions $c^{x}_j$, $c^{y}_j$, $c^{z}_j$, $d^{x}_j$, $d^{y}_j$, and $d^{z}_j$, with the constraint $D_j \equiv -i c^x_j c^y_j c^z_j d^x_j d^y_j d^z_j =1$ (namely, for any physical state $|\Phi \ra_{\rm phys}$, we require $D_j |\Phi \ra_{\rm phys} = |\Phi \ra_{\rm phys}$). Therefore, $\sigma_j^\alpha \tau_j^\beta |\Phi \ra_{\rm phys} = i c_j^\alpha d_j^\beta  |\Phi \ra_{\rm phys}$.
In terms of the Majoranas, the Hamiltonian can be rephrased as
\begin{eqnarray}
&& \mathcal{H}'_0 = i\sum_{\langle j k \rangle} \hat{u}_{jk} J_{jk} \sum_{\alpha=x,y,z}c^{\alpha}_{j}c^{\alpha}_{k},
\end{eqnarray}
and the $W_p$ terms are the same as in Eq.~(\ref{W-terms}) with $\hat{u}_{jk} \equiv -id^{\lambda}_j d^{\lambda}_k$ for $\lambda$-link $\la j k \ra$. 

For long-wavelength description, much of the development in Sec.~\ref{Sec:spinless_case} can be directly applied here with the replacement, $c\rightarrow c^\alpha$, $f\rightarrow f^\alpha$, $\alpha=x,y,z$. 
We now have three fermion species with identical dispersion taken to be similar to that in Fig.~\ref{2legladder_spec}, and we introduce right and left moving complex fermion fields $f^\alpha_{R/L}$ as in the spinless case.  Under SU(2) spin rotations, the triple $f^{x,y,z}$ transforms in the same way as the physical spin $\sigma^{x,y,z}$.

Just as in the MOL case in Sec~\ref{Sec:spinless_case}, we first establish the fixed point structure ignoring the gauge field fluctuations.
In order to study the stability of such gapless SU(2)-invariant Majorana spin liquid under weak perturbations, we write down most general four-fermion interactions and perform Renormalization Group (RG) studies. The allowed four-fermion interactions are highly constrained by symmetry. In addition to the SU(2) spin rotation invariance, these terms must be preserved by Projective Symmetry Group (PSG)\cite{WenPSG} of spatial translational symmetry, time reversal symmetry, and leg interchange symmetry. We list the symmetry transformations in Table~\ref{tab:transfprops} and write the allowed non-chiral interactions (i.e., connecting right and left movers) as
\begin{eqnarray}
\nonumber \mathcal{H}_{\rm int} &=& u_{\rho}\mathcal{J}_R \mathcal{J}_L - u_{\sigma 1}\vec{\mathcal{J}}_R \cdot \vec{\mathcal{J}}_L + u_{\sigma 2} I^{\dagger}_{RL} I_{RL}\\
&+& w_4 \left( I_{RL} I_{RL}+ \Hc \right),~\label{4ferm-int}
\end{eqnarray}
where we defined 
\begin{eqnarray}
&& \mathcal{J}_P = \sum_{\alpha} f^{\alpha\dagger}_P f^{\alpha}_P,~\label{Jrho}\\
&& \mathcal{J}^{\alpha}_P = -i \sum_{\beta, \gamma} \epsilon^{\alpha \beta \gamma} f^{\beta\dagger}_P f^{\gamma}_P,~\label{Jsigma}\\
&& I_{RL} = \sum_{\alpha} f^{\alpha}_R f^{\alpha}_L.~\label{Irl}
\end{eqnarray}
The general expression $\mathcal{H}_{\rm int}$ in Eq.~(\ref{4ferm-int}) contains familiar-looking four-fermion terms $f^{\alpha\dagger}_R f^{\beta\dagger}_L f^{\gamma}_R f^{\delta}_L$ that conserve fermion number, and also terms $f^{\alpha}_R f^{\beta}_R f^{\gamma}_L f^{\delta}_L$ that do not conserve the fermion number but are nevertheless allowed by all symmetries of the problem.  The less familiar terms need to be considered since the microscopic Majorana Hamiltonian does not have $U(1)$ particle conservation, which is a new feature in such Majorana liquids.

\begin{table}  
\caption{PSG transformation properties of the continuum fields under $T_x$ (spatial translation symmetry), 
$\Theta$ (time reversal transformation plus gauge transformation),\cite{Kitaev06} $M$ (leg interchange transformation plus gauge transformation). We also note that under spin rotation, $\vec{f}_P=(f^x_P,~f^y_P,~f^z_P)$ and $\vec{f}^{\dagger}_P=(f^{x\dagger}_P~f^{y\dagger}_P,~f^{z\dagger}_P)$ transform as 3-dimensional vectors. Note that below, $P=R/L$ and $\bar{P} = -P = L/R$}
\label{tab:transfprops} 
\begin{ruledtabular}  
\begin{tabular}{c | c | c | c }  
& $T_x$ & $\Theta$ & $M$\\ 
  \hline
  $f^\alpha_{P} \to$ &  $e^{i P k_F} f^{\alpha}_{P}$ & $f^\alpha_{\bar{P}};~i\rightarrow -i$ & $-i f^\alpha_{P} $\\
  \hline
   $f^{\alpha \dagger}_{P} \to$ &  $e^{-i P k_F} f^{\alpha \dagger}_{P}$ & $f^{\alpha\dagger}_{\bar{P}};~i\rightarrow -i$ & $i f^{\alpha\dagger}_{P} $\\ 
\end{tabular}  
\end{ruledtabular}  
\end{table}

We remark that the time reversal and translation symmetries alone would allow yet other terms expressed as $f^{\alpha\dagger}_R f^{\beta}_R f^{\gamma}_R f^{\delta}_L$ and in fact would also allow a bilinear term $(i I_{RL} + \Hc)$ in the Hamiltonian that would immediately open a gap in the spectrum.  However, these terms are prohibited if we also require the leg interchange symmetry, which is hence crucial for the time-reversal invariant SU(2) MSL. 
 
The weak-coupling differential RG equations are
\begin{eqnarray}
&& \dot{u}_{\rho} = \frac{1}{2\pi v}\left( u_{\sigma 2}^2 + 2 u_{\sigma 1} u_{\sigma 2} - 4 w_4^2 \right),~\label{urho}\\
&& \dot{u}_{\sigma 1} = \frac{1}{2\pi v} \left(- u_{\sigma 1}^2 + 2 u_{\sigma 1} u_{\sigma 2}\right),~\\
&& \dot{u}_{\sigma 2} = \frac{1}{2\pi v}\left( -3u_{\sigma 2}^2 - 6u_{\sigma 1} u_{\sigma 2} - 4 w_4^2 \right),~\\
&& \dot{w}_{4} = \frac{1}{2\pi v}  \left(-2u_{\sigma 1}-4u_{\sigma 2} - 4u_\rho \right)w_4 ,~\label{w4}
\end{eqnarray}
where $v$ is the Fermi velocity of right and left movers and $\dot{O}\equiv dO/d\ell$ with $\ell$ being logarithm of the length scale.  The only fixed points have  $u^*_{\sigma 1}=u^*_{\sigma 2} = w_4^* = 0$. Stability to small deviations in $w_4$ requires $u^*_\rho > 0$. If we consider small deviations in $u_{\sigma 1}$ and $u_{\sigma 2}$ setting $w^*_4=0$, the RG equations can be written as
\begin{eqnarray}
&& \dot{g}_{\rho} \equiv 3 \dot{u}_\rho + \dot{u}_{\sigma 2} = 0 ,~\\
&& \dot{u}_{\sigma 1} = \frac{1}{2\pi v} (-u^{2}_{\sigma 1 } + 2 u_{\sigma 1} u_{\sigma 2}),~\\
&& \dot{u}_{\sigma 2} = \frac{1}{2\pi v} (-3 u^{2}_{\sigma 2 } - 6 u_{\sigma 1} u_{\sigma 2}),~
\end{eqnarray}
and the last two equations are essentially identical to the RG equations in a level-one SU(3) Wess-Zumino-Witten (WZW) model discussed by Itoi and Kato.\cite{Itoi97}  Translated from their analysis, the stability to small deviation in $u_{\sigma 1}$ and $u_{\sigma 2}$ requires $u_{\sigma 1}>0,~u_{\sigma 1} + u_{\sigma 2}>0$. In a stable flow, $u_\rho$ reaches some fixed value, $u^*_\rho > 0$, and is strictly marginal; $u_{\sigma 1}$ and $u_{\sigma 2}$ approach zero from the specific region described above and are marginally irrelevant; finally, $w_4$ flows to zero as long as $u^*_\rho>0$ and is irrelevant. Thus, we have one Luttinger parameter in the ``charge'' sector. 
In Appendix~\ref{Appendix:SU(2)observables}, we give the fixed-point theory of the SU(2) MSL and list observables that can be obtained as fermion bilinears. We find that spin operator, Eq.~(\ref{spin}), spin-nematic operator, Eq.~(\ref{spin-nematic}), and bond-energy operator, Eq.~(\ref{bond-energy}), have correlations that decay in a power law with oscillations at incommensurate wave vectors, which is one of the hallmarks of such Majorana spin liquids as we discussed in Ref.~\onlinecite{Lai_MSL11} in a 2d example.

The inclusion of the $Z_2$ gauge field fluctuations in this quasi-1d gapless MSL can be discussed as in the spinless case (see also Appendix~\ref{Appendix:deconfinement}).  The space-time gauge field fluctuations are suppressed by the destructive interference arising from the incommensurate momenta carried by the fermion fields.  Thus, the system retains three gapless modes, but the local energy observable obtains new oscillating contributions.

We can also consider directly allowed perturbations going beyond the exactly solvable model.  For example, $\tau^x_i \tau^x_j$ terms on the vertical links $\la ij \ra$ can be expressed as a product of three $c$-fermion strings, one for each flavor, and will oscillate at wavevector $3k_F$ with power law $X^{-3/2}$ in the free parton case.  This is consistent with the schematic analysis in Appendix~\ref{Appendix:deconfinement} extended to multiple parton fields, where a vison can be seen as introducing a half-vortex for each flavor.  The described low-energy theory is hence stable to generic perturbations in the sense of retaining the gapless fields, while the local energy observable that is symmetric under the leg-interchange obtains additional contributions oscillating at $3k_F$ (which in turn induces new contributions to other observables as discussed in Appendix~\ref{Appendix:SU(2)observables}).

\section{Discussion}
\label{sec:concl}
Motivated by recent proposal of SU(2)-invariant Majorana Spin Liquids by Biswas\etal\cite{Biswas11} and the realization of the SU(2) MSL in an exactly solvable model,\cite{Fawang09, Yao10, Lai_MSL11} we studied the MOL and SU(2) MSL on the 2-leg ladder.  Perturbing away from the exactly solvable points, in the MOL, there is only a strictly marginal four-fermion interaction and hence it is stable to residual interactions. In the SU(2) MSL, there are several allowed four-fermion terms, but it is stable against these in a large parameter regime. Furthermore, we also show that such gapless Majorana liquids persist against $Z_2$ gauge field fluctuations. Some time ago, Shastry and Sen\cite{Shastry97} studied an SU(2) MSL for a 1d Heisenberg chain at mean field level.  Our description of the microscopically realized quasi-1d SU(2) MSL can be viewed as providing a theory beyond mean field for more general such states and distinguishes them from the Bethe phase of the 1d Heisenberg chain.
The stable MOL and SU(2) MSL phases that we find are new quasi-1d phases, and we suggest numerical studies such as Density Matrix Renormalization Group (DMRG)\cite{Jiang11} to test our theoretical ideas of their stability. The DMRG studies can also determine the Luttinger parameters of the fixed-point MOL and SU(2) MSL theories.

The presence of gapless matter fields is the key against confining effects of $Z_2$ gauge field fluctuations in (1+1)D, see Appendix~\ref{Appendix:deconfinement}. Without such gapless matter, the gapped phases realized in Kitaev-type models on 2-leg ladders in our model are likely unstable to general generic perturbations, and this prediction can be checked by DMRG studies. This is reminiscent of a picture where gapless matter fields can suppress monopoles in a (2+1)D compact electrodynamics and thus make gapless U(1) spin liquids with sufficiently many Dirac points or with Fermi surfaces stable,\cite{LeeNagaosaWen, Rantner2002, Hermele2004} while gapped U(1) spin liquids would be unstable to confinement in (2+1)D. An interesting finding is that allowing $Z_2$ gauge fluctuations in our quasi-1d Majorana liquids leads to new contributions to various observables, with different characteristic wavevectors and potentially slower power laws compared to the mean field, cf.\ Appendix~\ref{Appendix:SU(2)observables}.

Let us discuss possible extensions of this work.  Throughout, we focused on the MSL phase in which all couplings of the residual interactions, Eq.~(\ref{4ferm-int}), converge to finite fixed point values in RG thinking.  In principle, one can analyze situations where some of the residual interactions are relevant and explore possible nearby phases and characterize their properties using the observables listed in Appendix~\ref{Appendix:SU(2)observables}.  Such theoretical analysis combined with DMRG studies\cite{Sheng09} can give a complete phase diagram.

As discussed in Biswas\etal\cite{Biswas11} and in our earlier work,\cite{Lai_MSL11} the effects of Zeeman field on the SU(2) MSL are interesting. The Zeeman magnetic field only couples to $f^x$ and $f^y$ fermions, and we can define $f^\dagger_{\pm}=(f^{x\dagger} \pm i f^{y\dagger})/\sqrt{2}$ which carry $S^z = \pm 1$, while $f^{z\dagger}$ carries $S^z=0$ and remains unaltered. In the presence of the Zeeman field, the spin SU(2) rotation symmetry is broken and only $S^z$ is conserved. In Appendix~\ref{Appendix:Zeeman} we write down general four-fermion interactions based on symmetry arguments and perform weak coupling RG analysis.  Our RG equations~(\ref{w+-})-(\ref{lambdaz}) interestingly show that instabilities only occur in the $f^{\pm}$ channel but not in the $f^z$ channel.  Hence, the $f^z$ partons are always gapless no matter how large the field is and can give metal-like contribution to specific heat and thermal conductivity, which is qualitatively similar to what we found previously in our 2d MSL model.\cite{Lai_MSL11} 

Last but not least, it is intriguing to understand how the ladder descendants of the MOL and SU(2) MSL relate to the mother 2d phases.  A systematic way to access these could be via increasing the number of legs. It seems difficult to increase the number of legs in our toy 2-leg square ladder model while maintaining the spin SU(2) symmetry of the MSL, but actually it can be achieved if we consider decorated square ladder.\cite{Baskaran09, Lai_MSL11}  
One more interesting direction is to consider new types of SU(2)-invariant spin liquid wave functions motivated by the Kitaev-like SU(2) MSL writing of the spin operators and search for more realistic models in 1d and 2d that may harbor such states.

\appendix
\section{Stability of gapless U(1) matter against $Z_2$ gauge field fluctuations in (1+1)D}\label{Appendix:deconfinement}
We need to address the issue whether the gapless parton field picture is stable against allowing $Z_2$ gauge field fluctuations.  It is well-known that the simplest so-called even $Z_2$ gauge theory is confining in (1+1)D; this persists also in the presence of gapped matter fields, and quasi-1d Kitaev-type models with gapped partons would suffer from this instability.  We will argue, however, that gapless parton fields can eliminate this instability, particularly when they carry incommensurate momenta.

We first give a heuristic argument.  Let us consider the simplest model of a $Z_2$ gauge field coupled to a U(1) matter field, with (1+1)D action
\begin{eqnarray}
\mathcal{S} = -\beta \sum_{\la j k \ra} \sigma_{jk}\cos{(\phi_j - \phi_k)} - K \sum_{\square} \sigma_{12} \sigma_{23} \sigma_{34} \sigma_{41}.~~~~~
\end{eqnarray}
For $K \to \infty$, we choose the gauge $\sigma_{jk}=1$ and obtain an XY model in the $\phi$ variables.  There is a Kosterlitz-Thouless transition at some critical $\beta_c$ and gapless phase for $\beta > \beta_c$.  Now, let us consider large $K$ and large $\beta$ limit.  Starting with no $Z_2$ fluxes and no vortices, since both $\sigma$ and $\phi$ are almost fixed, the insertion of a $Z_2$ flux (``vison'') can be treated as creating a $\pi$-vortex in the $\phi$.
Explicitly, we can rewrite $\sigma_{jk} \cos{(\phi_j - \phi_k)} = \cos{[\phi_j - \phi_k -\pi(1-\sigma_{jk})/2]}$.  The vison insertion can be carried out by changing $\sigma_{jk}$ from $1$ to $-1$ on a cut from infinity to the vison location.  This is a $\pi$-phase cut for the $\phi$ variables and can be best accomodated by a gradual winding by $\pi$ as we go around the vison from one side of the cut to the other; hence, we get a half-vortex in the $\phi$.  We expect that for sufficiently large $\beta$, the half-vortex insertions are irrelevant because of their high energy cost, which means we have a phase without proliferation of half-vortices, and then we do not need to worry about the dynamics of the $Z_2$ gauge field which could potentially produce confinement.

Thus, it is possible to avoid confinement of (1+1)D $Z_2$ gauge fields if we have gapless matter field.  For several gapless matter fields, there is a proportional increase in the energy cost of the vison insertion and hence its irrelevance.  The above argument is valid for matter fields at integer filling.  It is well-known that vortices in (1+1)D U(1) systems can be further suppressed if the matter field is at non-integer filling due to Berry phase effects, and such a suppression is complete for incommensurate matter density.  Heuristically, we expect the vison insertions to obtain similar Berry phases as half-vortices and hence to also experience complete suppression at incommensurate density.  We present a more formal derivation\cite{SenthilFisher_Z2} tailored to our needs below.

We consider a general $Z_2$ gauge theory plus $U(1)$ matter field (represented by quantum rotors) on a $d$-dimensional cubic lattice with a Hamiltonian\cite{SenthilFisher_Z2}
\begin{eqnarray}
\nonumber \mathcal{H}&=&-t\sum_{\la r r'\ra} \hat{\sigma}_{rr'}^z\cos{(\hat{\phi}_r-\hat{\phi}_{r'})} + \frac{U}{2}\sum_r (\hat{n}_r -\bar{n})^2\\
&& -K\sum_{\square}\hat{\sigma}^z_{12}\hat{\sigma}^z_{23}\hat{\sigma}^z_{34}\hat{\sigma}^z_{41} - \Gamma \sum_{\la rr' \ra}\hat{\sigma}^x_{rr'} ~,
\label{HU1Z2}
\end{eqnarray}
where $\hat{n}_r$ is the number operator conjugate to the phase $\hat\phi_r$ at site $r$ and $\bar{n}$ is the average density.  The Hilbert space constraint is 
\begin{eqnarray}
e^{i\pi \hat{n}_r} \prod _{r' \in r} \hat{\sigma}^x_{rr'} = 1 ~.
\end{eqnarray}
We proceed to treat the system using standard Euclidean path integral formalism in the $\sigma^z$-$\phi$ basis.  We implement the constraint at each site $r$ and temporal coordinate $\tau$ by using the identity
\begin{eqnarray*}
\nonumber \delta_{e^{i\pi n_{r}} \cdot \prod_{r'\in r}\sigma^{x}_{rr'}=1}=\frac{1}{2}\sum_{\lambda(r,\tau)=\pm1}e^{i\pi \frac{1-\lambda}{2}(n_r+\sum_{r'\in r}\frac{1-\sigma^x_{rr'}}{2})} ~.
\end{eqnarray*}
After standard development of the path integral for the Ising gauge fields, we can write the partition function as
\begin{widetext}
\begin{eqnarray}
\nonumber \mathcal{Z}&=& \sum_{\{S^z_{rr'}(\tau);\lambda(r,\tau)\}} \int_0^{2\pi}\mathcal{D}\phi_{r}(\tau) \sum_{\{n_r(\tau)\}} e^{\sum_{P} K_{P}S^z_{12}S^z_{23}S^z_{34}S^z_{41}} \times e^{t \delta\tau \sum_{\tau, \la r r' \ra} S^{z}_{rr'}(\tau)\cos{[\phi_{r}(\tau)-\phi_{r'}(\tau)}]}\times\\
&&  \hspace{6cm} \times e^{-\frac{U \delta \tau}{2}\sum_{\tau, r} [n_r(\tau) -\bar{n}]^2 + i \sum_{\tau, r} n_r(\tau) [\phi_r(\tau + \delta\tau) - \phi_r(\tau) + \pi \frac{1 - \lambda(r,\tau)}{2}]} ~.
\end{eqnarray}
\end{widetext}
Here we used $S^z_{rr'}$ to denote eigenvalues of $\hat{\sigma}^z_{rr'}$ on the spatial links and elevated the auxiliary fields $\lambda(r,\tau)$ to become Ising gauge fields on the temporal links, $S^z_{(r,\tau); (r,\tau+\delta\tau)} \equiv \lambda(r, \tau)$ (we use either field notation where more convenient);
$\sum_P$ is over all spatial and temporal plackets, $K_P = \{ K_{\rm spat}, K_{\tau}\}$, with $K_{\rm spat} = K \delta\tau$ and $\tanh K_{\tau} = e^{-2\Gamma \delta\tau}$. 

Now we can use a variant of XY duality transformation\cite{Polyakov,Peskin,Savit} to go from the $\phi$ and $n$ variables to real-valued ``currents'' $\vec{j}_{\rm spat} = \{ j_{r, r+\hat{e}_1}, j_{r, r+\hat{e}_2}, \dots, j_{r, r+\hat{e}_d} \}$ (where $\hat{e}_{k=1\dots d}$ represent unit lattice vectors) and $j_{\tau}$ appearing as follows:
\begin{widetext}
\begin{eqnarray}
\nonumber 
&& e^{t \delta\tau S^z_{rr'}(\tau)\cos [\phi_r(\tau)-\phi_{r'}(\tau)]} \simeq \sum_{p_{rr'}(\tau)=-\infty}^{+\infty} e^{-\frac{t \delta\tau}{2}\left[\phi_{r'}(\tau)-\phi_{r}(\tau) + \pi\frac{1-S^z_{rr'}(\tau)}{2} - 2\pi p_{rr'}(\tau)\right]^2} \\
&&\hspace{4cm} = \sum_{p_{rr'}(\tau)=-\infty}^{+\infty} \int_{-\infty}^{+\infty} dj_{rr'}(\tau) e^{-\frac{j_{rr'}^2(\tau)}{2t \delta \tau} + i j_{rr'}(\tau)\left[\phi_{r'}(\tau)-\phi_{r}(\tau) + \pi \frac{1-S^z_{rr'}(\tau)}{2} - 2\pi p_{rr'}(\tau)\right]} ~,\\
&& \sum_{n_r(\tau)=-\infty}^{+\infty} F[n_{r}(\tau)] = \int_{-\infty}^{+\infty} dj_{\tau}(r,\tau)\sum_{p_\tau(r,\tau)=-\infty}^{+\infty} e^{-i j_{\tau}(r,\tau) \cdot 2\pi p_{\tau}(r,\tau)} F[j_\tau(r,\tau)] ~.
\end{eqnarray}
\end{widetext}
In the first line, we approximated the left hand side by a standard Villain form; we also dropped constant numerical factors throughout.
For short-hand, we write space-time points as $i=(r,\tau)$ and define space-time vector $p_{i,\mu=1\dots d+1} = \{ \vec{p}_{\rm spat}, p_\tau \}$, with $\vec{p}_{\rm spat} = \{ p_{r, r+\hat{e}_1}, p_{r, r+\hat{e}_2}, \dots, p_{r, r+\hat{e}_d} \}$.
Then we can divide configurations $\{ p_{i\mu} \}$ into classes $C_p$ equivalent under integer-valued gauge transformations $p_{i\mu} \to p_{i\mu} + \nabla_{\mu} N_i$ and perform the configuration summation as
\begin{eqnarray*}
\sum_{\{ p_{i\mu}\}=-\infty}^{+\infty} F[\{ p_{i\mu}\}] = \sum_{C_P} \sum_{N_i =-\infty}^{\infty} F[\{ p_{i\mu} = p_{i\mu}^{(0)} + \nabla_{\mu} N_i\}] ~,
\end{eqnarray*}
where $p_{i\mu}^{(0)}$ is one representative of a class; the results do not depend on the specific choices of $p^{(0)}$ but only on the ``vorticities'' $q_{\mu\nu} = \nabla_\mu p_\nu - \nabla_\nu p_\mu$ characterizing the classes.  Using the $N_i$ variables, we can extend the $\phi_i$ integrations to $(-\infty, +\infty)$ and obtain
\begin{widetext}
\begin{eqnarray}
\nonumber \mathcal{Z} = \sum_{\{S^z_{rr'}(\tau); \lambda(r,\tau)\}} \sum_{C_P}&&  \int_{-\infty}^{\infty} \mathcal{D}\vec{j}_{\rm spat} \mathcal{D}j_{\tau} ~ \delta(\vec{\nabla} \cdot \vec{j}_{\rm spat} + \nabla_{\tau} j_{\tau} = 0)\times 
e^{\sum_P K_P S^z_{12}S^z_{23}S^z_{34}S^z_{41}} \times \\
\nonumber && \times e^{-\sum_{\tau, \la rr'\ra} \frac{j_{rr'}(\tau)^2}{2t\delta \tau} + i\sum_{\tau, \la rr' \ra} j_{rr'}(\tau) \left[\pi\frac{1-S^z_{rr'}(\tau)}{2} - 2\pi p^{(0)}_{rr'}(\tau)\right]} \times \\
 &&  \times e^{-\sum_{\tau, r} \frac{U\delta \tau}{2} [j_\tau(r,\tau) -\bar{n}]^2 + i\sum_{\tau,r} j_{\tau}(r,\tau) \left[\pi\frac{1-\lambda(r,\tau)}{2} - 2\pi p^{(0)}_\tau(r,\tau)\right]} ~.
\end{eqnarray}
\end{widetext}

The above result holds in general (d+1)D,\cite{SenthilFisher_Z2} and from now on we specialize to (1+1)D system.  We solve the current conservation condition by writing $j_{\tau} = \bar{n} + \frac{\nabla_x \theta}{\pi} = \frac{\nabla_x (\theta + \bar{\theta})}{\pi}$, with $\bar{\theta}(x,\tau) \equiv \pi \bar{n} x$, $x$ being the spatial coordinate on the dual lattice, and $j_{x} = -\frac{\nabla_\tau \theta}{\pi} = -\frac{\nabla_\tau (\theta + \bar{\theta})}{\pi}$.  The dual field $\theta$ encodes coarse-grained fluctuations in the particle number.

We have only temporal plackets, on which we define ``vorticity'' $q = \vec{\nabla} \times \vec{p} = \nabla_x p_\tau - \nabla_\tau p_x$ and ``vison number'' $n^{\rm vison} = \vec{\nabla} \times (1 - \vec{S^z})/2 \mod 2 = 0$ or $1$ corresponding to $S^z_{12} S^z_{23} S^z_{34} S^z_{41} = 1$ or $-1$.  We can absorb any modulo 2 shifts from $n^{\rm vison}$ by redefining $q$ and write the partition function as
\begin{widetext}
\begin{eqnarray}
\mathcal{Z} = \sum_{\vec{S^z}} \sum_{q} \int_{-\infty}^{\infty} \mathcal{D}\theta e^{\sum_P K_{P}(1 - 2 n^{\rm vison}_P)} \times e^{-\sum \frac{U\delta \tau}{2} \frac{(\nabla_x \theta)^2}{\pi^2} - \sum \frac{1}{2t\delta \tau} \frac{(\nabla_\tau \theta)^2}{\pi^2} + i \sum 2(\theta + \pi \bar{n} x)\times (q - \frac{1}{2} n^{\rm vison})} ~.
\end{eqnarray}
\end{widetext}

This is the main result, which we can now analyze in a number of standard ways.  We can integrate out the field $\theta$ and obtain a Coulomb gas representation.  In the absence of the $Z_2$ gauge field (e.g., $K \to \infty$ and $n^{\rm vison}=0$), we get familiar integer-valued charges $q$ representing vortices of the $U(1)$ matter system.  On the other hand, for any finite $K$ we get effectively half-integer charges $m = q - \frac{1}{2} n^{\rm vison} \in \frac{1}{2} \times \mathbb{Z}$ with only short-scale energetics difference between integer and half-integer charges.  We also see Berry phases $e^{i 2\pi\bar{n} x}$ for a vortex insertion in the presence of non-zero background density and halving of the Berry phase for a vison insertion.  Alternatively, we can consider postulating some local energetics penalty for large values of $m$ and perfom the summation over $m$ to obtain terms like 
\begin{equation}
\lambda_{1/2} \cos(\theta + \pi\bar{n} x) + \lambda_1 \cos(2\theta + 2\pi\bar{n} x) + \dots ~,
\end{equation}
where we ommitted possible phase shifts in the cosines for brevity. 
The $\lambda_1$ term is the familiar term in the dual sine-Gordon theory for a Luttinger liquid of bosons that represents allowing vortices, while the $\lambda_{1/2}$ term can be now interpreted as effectively allowing half-vortices if the matter is coupled to $Z_2$ gauge fields.  Crucially, both vortices and visons experience destructive interference effects for incommensurate $\bar{n}$.  On the other hand, for commensurate $\bar{n}$ the vison insertions can still be rendered irrelevant by going deep enough into the Luttinger phase or increasing the number of gapless fields as discussed below.

We can generalize the above result to the case with several matter fields $\phi_\alpha$ coupled to the same $Z_2$ gauge field by replacing the Berry phase $2(\theta + \pi \bar{n} x)\times (q - \frac{1}{2} n^{\rm vison})$ with $\sum_\alpha 2(\theta_\alpha + \pi \bar{n}_\alpha x)\times (q_\alpha - \frac{1}{2} n^{\rm vison})$.  Here the summation over vison numbers leads effectively to terms like $\lambda_{1/2} \cos(\sum_\alpha \theta_\alpha + \pi \sum_\alpha \bar{n}_\alpha x)$.  We can see that for three identical flavors with incommensurate $\bar{n}$ as happens in the SU(2)-invariant MSL, the destructive interference effects will wash out any vison insertions (including any combinations with non-vison terms).

Looking back at the one-component case, we could rationalize the above structure more quickly by thinking about the theory Eq.~(\ref{HU1Z2}) as coming from a formal splitting of some physical boson field $e^{i \phi_{\rm phys}}$ into two halves:\cite{SenthilFisher_Z2} schematically, $e^{i \phi_{\rm phys}} = e^{i 2\phi}$.  Then the described gapless phase can be thought of as a (1+1)D analogue of the ``Higgs phase'' that is expected\cite{SenthilFisher_Z2} to reproduce the conventional ``superfluid'' (here, quasi-long-range ordered) phase of the physical bosons.  Indeed, in the derived harmonic liquid description in terms of the dual field $\theta$, we can change to new variable $\theta_{\rm phys} = \theta/2$ canonically dual to $\phi_{\rm phys}$ and note that the identified vison insertion operator $e^{i \theta} = e^{i 2\theta_{\rm phys}}$ is the same as the conventional vortex insertion in $\phi_{\rm phys}$.  We still like to show the above more formal derivation as it is not tied to the specific origin of the parton field $\phi$.  For example, in Sec.~\ref{MOL:observable} the conjugate pair $\{\phi, \theta\}$ arose from bosonizing the long-wavelength fermionic parton Hamiltonian, and we can continue using these fields in calculations but remember to include the $Z_2$ gauge fluctuation effects by allowing local energy terms like $\lambda_{1/2} \cos(\theta + \pi\bar{n} x)$.  The same formal treatment also holds transparently for the multi-flavor generalization where the parton fields provide a very convenient description of the unconventional gapless phase, which has the same number of gapless modes as in the parton mean-field, but with the identified new contributions to the local energy once we go beyond the mean field and include $Z_2$ gauge field fluctuations.

\section{Fixed-point theory and observables in the SU(2) Majorana spin liquid}\label{Appendix:SU(2)observables}
We use Bosonization to re-express the low energy fermion operators, 
\begin{eqnarray}
f^{\alpha}_{P} = \eta_\alpha e^{i (\varphi_{\alpha}+P\theta_{\alpha})},~
\end{eqnarray}
with canonical conjugate boson fields:
\begin{eqnarray}
&& \left[\varphi_{\alpha}(x), \varphi_{\beta}(x')\right] = \left[\theta_{\alpha}(x), \theta_{\beta}(x')\right] = 0,~\label{comut1}\\
&& \left[ \varphi_{\alpha}(x), \theta_{\beta}(x')\right] = i\pi \delta_{\alpha \beta}\Theta(x-x'),~\label{comut2}
\end{eqnarray}
where $\Theta(x)$ is the Heaviside step function and we have introduced Klein factors, the Majorana fermions with $\{ \eta_{\alpha},\eta_\beta\}=2\delta_{\alpha \beta}$, which assure that the fermion fields with different flavors anti-commute with one another.

According to the RG analysis in Sec.~\ref{Sec:SU(2)_case}, at the fixed point of the stable SU(2) MSL phase, only the coupling $u_{\rho}$ is strictly marginal and will renormalize the Luttinger parameter $g$ in the ``charge" sector.  The effective bosonized Lagrangian is
\begin{eqnarray}
\nonumber \mathcal{L}^{SU(2)}_{MSL}=&&\frac{1}{2\pi g}\left[\frac{1}{v_\rho}(\partial_\tau \theta_{\rho})^2+v_{\rho}(\partial_x \theta_{\rho})^2\right]\\
 && +\sum_{\mu=1,2}\frac{1}{2\pi}\left[ \frac{1}{v}(\partial_\tau \theta_{\mu})^2+v (\partial_x \theta_{\mu})^2\right],~\label{fixed-point-action}
 \end{eqnarray}
where we defined
\begin{eqnarray}\label{Def: SU(2) bosons}
&& \theta_{\rho}=\frac{1}{\sqrt{3}}\left(\theta_x + \theta_y +\theta_z\right),~\\
&& \theta_{1}=\frac{1}{\sqrt{2}}\left(\theta_x -\theta_y \right),~\\
&& \theta_{2}=\frac{1}{\sqrt{6}}\left(\theta_x+\theta_y - 2\theta_z \right),~
\end{eqnarray}
and similarly for the $\varphi$-s, which preserves the commutation relations, Eqs.~(\ref{comut1})-(\ref{comut2}). Stability against the $w_4$ term in Eq.~(\ref{4ferm-int}) requires $g\leq 1$. 

For the observables characterizing the SU(2) MSL phase, as discussed in Ref.~\onlinecite{Lai_MSL11}, we can use spin operators, 
\begin{eqnarray}\label{spin}
\vec{S}_j = \frac{\vec{\sigma}_j}{2},~
\end{eqnarray}
 bond energy operators,
\begin{eqnarray} \label{bond-energy}
\mathcal{B}_{jk} = i u_{jk} J_{jk} \sum_{\alpha} c^{\alpha}_j c^{\alpha}_k,~
\end{eqnarray}
and spin-nematic operators
\begin{eqnarray}\label{spin-nematic}
P^{+}_{jk}=S^{+}_jS^{+}_k .
\end{eqnarray}
The latter can be related to the usual traceless rank two quadrupolar tensor defined as 
\begin{eqnarray}
\mathcal{Q}^{\alpha \beta}_{jk}=\frac{1}{2}\left( S^{\alpha}_j S^\beta_k +S^\beta_j S^\alpha_k \right)-\frac{1}{3}\delta^{\alpha \beta}\la \vec{S}_j \cdot \vec{S}_k \ra,~
\end{eqnarray}
through $P^{+}_{jk} = \mathcal{Q}^{xx}_{jk}-\mathcal{Q}^{yy}_{jk}+2 i \mathcal{Q}^{xy}_{jk}$.

We expand the observables in terms of the continuum complex fermion fields and organize according to the momentum and the leg interchange symmetry, i.e.\ symmetric ($s$) or anti-symmetric ($a$) under the leg interchange:
\begin{eqnarray}
&&S^{\alpha,s}_{Q=0}=-i\sum_{\beta, \gamma} \epsilon^{\alpha \beta \gamma}(f^{\beta\dagger}_R f^\gamma_R+f^{\beta\dagger}_L f^\gamma_L),~\\
&&\mathcal{B}^{s}_{Q=0} = \sum_{\beta}(f^{\beta\dagger}_R f^\beta_R + f^{\beta\dagger}_L f^\beta_L), ~\\
&&\mathcal{Q}^{\alpha \alpha, s}_{Q=0} = \sum_{\beta \neq \alpha}(f^{\beta\dagger}_R f^\beta_R + f^{\beta\dagger}_L f^\beta_L), ~\\
&&\mathcal{Q}^{\alpha \neq \beta, s}_{Q=0} = \sum_{P=R/L} (f^{\alpha\dagger}_P f^\beta_P + f^{\beta\dagger}_P f^\alpha_P), ~\\
&& S^{\alpha,a}_{k_{FR}+k_{FL}}=-i \sum_{\beta,\gamma} \epsilon^{\alpha \beta \gamma} f^\beta_R f^\gamma_L,~\label{TRB_SR+L}\\
&& \mathcal{B}^{a}_{k_{FR}+k_{FL}} = -i\sum_{\beta} f^{\beta}_R f^{\beta}_L,~\\
&& \mathcal{Q}^{\alpha \alpha,a}_{k_{FR}+k_{FL}}=-i \sum_{\beta \not= \alpha}f^{\beta}_R f^\beta_L,~\\
&&\mathcal{Q}^{\alpha \neq \beta, a}_{k_{FR}+k_{FL}} = -i\left( f^{\alpha}_R f^{\beta}_L+f^{\beta}_R f^{\alpha}_L\right),~\label{TRB_QR+L}\\
&& S^{\alpha,s}_{k_{FR}-k_{FL}}= -i \sum_{\beta,\gamma} \epsilon^{\alpha \beta \gamma} f^{\beta \dagger}_L f^\gamma_R,~\\
&&\mathcal{B}^{s}_{k_{FR}-k_{FL}}=\sum_\beta f^{\beta\dagger}_{L}f^{\beta}_R,~\\
&&\mathcal{Q}^{\alpha \alpha, s}_{k_{FR}-k_{FL}}=\sum_{\beta \not= \alpha}f^{\beta\dagger}_L f^\beta_R,~\\
&&\mathcal{Q}^{\alpha \neq \beta, s}_{k_{FR}-k_{FL}}=f^{\alpha\dagger}_L f^\beta_R + f^{\beta \dagger}_L f^\alpha_R,~\\
&& S^{\alpha,a}_{2k_{FP}} = -i \sum_{\beta, \gamma} \epsilon^{\alpha \beta \gamma} f^\beta_P  f^\gamma_P,~\label{TRB_2kf}
\end{eqnarray}
with $S^{\alpha}_{-Q}=S^{\alpha \dagger}_Q$, etc., and $O^{s/a}$ observables mean symmetric or anti-symmetric under the leg interchange. If the TRS is broken explicitly as in Appendix~\ref{Appendix:TRB}, all the above momenta are distinct. With TRS, $k_{FL}=-k_{FR}$, we have coincident momenta $k_{FR}+k_{FL}=0$ and $k_{FR}-k_{FL}=2 k_{FR}=-2k_{FL}$. Strictly speaking, with TRS, we should define $O^{a}_{Q=0}=O^{a}_{k_{FR}+k_{FL}}+\Hc$, instead of Eqs.~(\ref{TRB_SR+L})-(\ref{TRB_QR+L}); similarly, instead of Eq.~(\ref{TRB_2kf}), we should define $S^{\alpha,a}_{2 k_F}=S^{\alpha,a}_{2k_{FR}} + S^{\alpha,a}_{-2k_{FL}}$. In the present case, the listed terms with such equal momenta transform differently under leg interchange, which is encoded in the above definitions.   

The bosonized forms at $Q=0$ are:
\begin{eqnarray}
&& S^{x,s}_{Q=0}=4i \eta_z \eta_y \cos\bigg{(}\frac{\sqrt{3}\varphi_2-\varphi_1}{\sqrt{2}}\bigg{)} \cos \bigg{(}\frac{\sqrt{3}\theta_2-\theta_1}{\sqrt{2}}\bigg{)},~~~~~~~~\\
&& S^{y,s}_{Q=0}=4 i \eta_x\eta_z \cos\bigg{(}\frac{\sqrt{3}\varphi_2 + \varphi_1}{\sqrt{2}}\bigg{)}\cos\bigg{(}\frac{\sqrt{3}\theta_2 + \theta_1}{\sqrt{2}}\bigg{)},~~~~~~~~\\
&& S^{z,s}_{Q=0}=4i\eta_y\eta_x\cos\big{(}\sqrt{2}\varphi_{1}\big{)}\cos\bigg{(}\sqrt{2}\theta_1 \bigg{)},~~~~~~~~\\
&& \mathcal{B}^s_{Q=0}=\frac{\sqrt{3}}{\pi}\partial_x \theta_\rho,~~~~~~~~\\
&& \mathcal{Q}^{xx,s}_{Q=0}= -\frac{\partial_x \theta_1}{\sqrt{2} \pi} -\frac{\partial_{x}\theta_2}{\sqrt{6}\pi} ,~~~~~~~~\\
&& \mathcal{Q}^{yy,s}_{Q=0} = \frac{\partial_x \theta_1}{\sqrt{2} \pi} - \frac{\partial_x \theta_2}{\sqrt{6}\pi},~~~~~~~~\\
&& \mathcal{Q}^{zz,s}_{Q=0} = \frac{1}{\pi}\sqrt{\frac{2}{3}}\partial_x\theta_2,~~~~~~~~\\
&& \mathcal{Q}^{xy,s}_{Q=0} = 4 i \eta_y \eta_x \cos\left( \sqrt{2}\theta_1 \right) \sin\left( \sqrt{2}\varphi_1 \right),~~~~~~~~\\
&& \mathcal{Q}^{yz,s}_{Q=0} = 4 i \eta_z \eta_y \cos \bigg{(} \frac{ \sqrt{3} \theta_2 -\theta_1}{\sqrt{2}} \bigg{)} \sin \bigg{(}\frac{\sqrt{3}\varphi_2 - \varphi_1}{\sqrt{2}} \bigg{)},~~~~~~~~\\
&& \mathcal{Q}^{xz,s}_{Q=0} = 4 i\eta_z \eta_x \cos \bigg{(} \frac{\sqrt{3} \theta_2 + \theta_1}{\sqrt{2}} \bigg{)} \sin \bigg{(} \frac{\sqrt{3}\varphi_2 + \varphi_1}{\sqrt{2}} \bigg{)}.~~~~~~~~~
\end{eqnarray}
The corresponding scaling dimension in the fixed-point theory Eq.~(\ref{fixed-point-action}) is
\begin{eqnarray}
\Delta[\vec{S}^s_{Q=0}]=\Delta[\mathcal{B}^s_{Q=0}]=\Delta[\mathcal{Q}^{\alpha\beta,s}_{Q=0}]=1,~
\end{eqnarray}
which is not modified by the strictly marginal interactions. 

The bosonized forms at $Q_{+}\equiv k_{FR}+k_{FL}$ are:
\begin{eqnarray}
&& S^{x,a}_{Q_{+}} = 2 i \eta_z \eta_y e^{i (\frac{2}{\sqrt{3}} \varphi_\rho - \frac{\varphi_2}{\sqrt{6}} - \frac{\varphi_1}{\sqrt{2}})} \cos \bigg{(}\frac{\sqrt{3}\theta_2-\theta_1}{\sqrt{2}}\bigg{)},~~~~~~~~\\
&&  S^{y,a}_{Q_{+}} = 2i \eta_x \eta_z e^{i (\frac{2}{\sqrt{3}} \varphi_\rho -\frac{\varphi_2}{\sqrt{6}}+\frac{\varphi_1}{\sqrt{2}})}\cos \bigg{(} \frac{\sqrt{3}\theta_2+\theta_1}{\sqrt{2}}\bigg{)},~~~~~~~~\\
&& S^{z,a}_{Q_{+}}= 2i \eta_y \eta_x e^{i(\frac{2}{\sqrt{3}}\varphi_{\rho}+\sqrt{\frac{2}{3}}\varphi_{2})}\cos(\sqrt{2}\theta_{1}),~~~~~~~\\
&&  \mathcal{B}^{a}_{Q_{+}}=e^{i\frac{2}{\sqrt{3}}\varphi_\rho}\bigg{[} 2 e^{i\sqrt{\frac{2}{3}}\varphi_2} \cos(\sqrt{2}\varphi_1) + e^{-i 2\sqrt{\frac{2}{3}}\varphi_2} \bigg{]},~~~~~~~~\\
&& \mathcal{Q}^{xx,a}_{Q_{+}} =  e^{i\frac{2}{\sqrt{3}}\varphi_\rho} \bigg{[} e^{i(\sqrt{\frac{2}{3}}\varphi_2 - \sqrt{2} \varphi_1)}+e^{-i2\sqrt{\frac{2}{3}}\varphi_2}\bigg{]},~~~~~~~~\\
&&  \mathcal{Q}^{yy,a}_{Q_{+}} =  e^{i \frac{2}{\sqrt{3}}\varphi_\rho} \bigg{[} e^{i (\sqrt{\frac{2}{3}}\varphi_2 - \sqrt{2}\varphi_1 )}+e^{-i2 \sqrt{\frac{2}{3}}\varphi_2}\bigg{]},~~~~~~~~\\
&& \mathcal{Q}^{zz,a}_{Q_{+}} = 2 e^{i(\frac{2}{\sqrt{3}}\varphi_\rho  +\sqrt{\frac{2}{3}}\varphi_2 )} \cos (\sqrt{2}\varphi_1),~~~~~~~~\\
&& \mathcal{Q}^{xy,a}_{Q_{+}} = 2 \eta_x \eta_y e^{i(\frac{2}{\sqrt{3}}\varphi_\rho + \sqrt{\frac{2}{3}}\varphi_2 )}\sin (\sqrt{2}\theta_1),~~~~~~~~\\
&&  \mathcal{Q}^{yz,a}_{Q_{+}} = 2 \eta_y \eta_z e^{i(\frac{2}{\sqrt{3}}\varphi_\rho -\frac{\varphi_2}{\sqrt{6}}-\frac{\varphi_1}{\sqrt{2}})}\sin\bigg{(}\frac{\sqrt{3}\theta_2-\theta_1}{\sqrt{2}}\bigg{)},~~~~~~~~\\
&&  \mathcal{Q}^{xz,a}_{Q_{+}} = 2 \eta_x \eta_z e^{i(\frac{2}{\sqrt{3}}\varphi_\rho - \frac{\varphi_2}{\sqrt{6}}+\frac{\varphi_1}{\sqrt{2}})}\sin \bigg{(}\frac{\sqrt{3}\theta_2+\theta_1}{\sqrt{2}}\bigg{)}.~
\end{eqnarray}
The corresponding scaling dimension is
\begin{eqnarray}
\Delta[\vec{S}^{a}_{Q_{+}}] = \Delta [ \mathcal{B}^{a}_{Q_{+}}] &=& \Delta[\mathcal{Q}^{\alpha \beta,a}_{Q_{+}}]=\frac{2}{3}+\frac{1}{3g}.~
\end{eqnarray}
The bosonized forms at $Q_{-}\equiv k_{FR}-k_{FL}$ are:
\begin{eqnarray}
 &&  S^{x,s}_{Q_-} = 2i \eta_z \eta_y e^{i(\frac{2}{\sqrt{3}}\theta_\rho-\frac{\theta_2}{\sqrt{6}}-\frac{\theta_1}{\sqrt{2}})}\cos\bigg{(}\frac{\sqrt{3}\varphi_2-\varphi_1}{\sqrt{2}}\bigg{)},~~~~~~~ \\
&& S^{y,s}_{Q_{-}}= 2i \eta_x \eta_z e^{i (\frac{2}{\sqrt{3}}\theta_\rho -\frac{\theta_2}{\sqrt{6}}+\frac{\theta_1}{\sqrt{2}})} \cos\bigg{(}\frac{\sqrt{3}\varphi_2+\varphi_1}{\sqrt{2}}\bigg{)},~~~~~~~\\
&&  S^{z,s}_{Q_-}= 2i \eta_y \eta_x e^{i(\frac{2}{\sqrt{3}}\theta_\rho+\sqrt{\frac{2}{3}}\theta_2)}\cos(\sqrt{2}\varphi_1),~~~~~~~\\
 &&  \mathcal{B}^{s}_{Q_{-}} =  i e^{i\frac{2}{\sqrt{3}}\theta_\rho}\bigg{[} 2 e^{i \sqrt{\frac{2}{3}}\theta_2}\cos(\sqrt{2}\theta_1) + e^{-i2\sqrt{\frac{2}{3}}\theta_2} \bigg{]},~~~~~~~\\
 &&\mathcal{Q}^{xx,s}_{Q_{-}}=ie^{i\frac{2}{\sqrt{3}}\theta_\rho}\bigg{[}e^{i(\sqrt{\frac{2}{3}}\theta_2-\sqrt{2}\theta_1)}+ e^{-i 2 \sqrt{\frac{2}{3}}\theta_2}\bigg{]},~~~~~~~\\
 &&\mathcal{Q}^{yy,s}_{Q_{-}} = i e^{i \frac{2}{\sqrt{3}}\theta_\rho} \bigg{[}e^{i (\sqrt{\frac{2}{3}}\theta_2+\sqrt{2}\theta_1)} +e^{-i 2 \sqrt{\frac{2}{3}}\theta_2}\bigg{]},~~~~~~~\\
 &&  \mathcal{Q}^{zz,s}_{Q_{-}} = 2i e^{i (\frac{2}{\sqrt{3}}\theta_\rho + \sqrt{\frac{2}{3}}\theta_2)} \cos( \sqrt{2} \theta_1 ),~~~~~~~\\
 &&  \mathcal{Q}^{xy,s}_{Q_{-}}= 2i\eta_y \eta_x e^{i (\frac{2}{\sqrt{3}} \theta_{\rho} + \sqrt{\frac{2}{3}} \theta_2)} \sin(\sqrt{2} \varphi_1 ),~~~~~~~\\
 &&  \mathcal{Q}^{yz,s}_{Q_{-}} = 2i \eta_z \eta_y e^{i(\frac{2}{\sqrt{3}}\theta_\rho -\frac{\theta_2}{\sqrt{6}}-\frac{\theta_1}{\sqrt{2}})}\sin\bigg{(}\frac{\sqrt{3}\varphi_2-\varphi_1}{\sqrt{2}}\bigg{)},~~~~~~~\\
 &&  \mathcal{Q}^{xz,s}_{Q_{-}} = 2i \eta_z \eta_x e^{i (\frac{2}{\sqrt{3}} \theta_{\rho} - \frac{\theta_2}{\sqrt{6}} + \frac{\theta_1}{\sqrt{2}} )} \sin\bigg{(}\frac{\sqrt{3}\varphi_2+\varphi_1}{\sqrt{2}}\bigg{)}.~
\end{eqnarray}
The corresponding scaling dimension is
\begin{eqnarray}
 \Delta [ \vec{S}^{s}_{Q_{-}}] = \Delta [ \mathcal{B}^{s}_{Q_{-}} ]&=&\Delta [\mathcal{Q}^{\alpha \beta,s}_{Q_{-}}]=\frac{2}{3}+\frac{g}{3}.~
\end{eqnarray}
The bosonized forms at the $2k_{FP}$ are:
\begin{eqnarray}
&& S^{x,a}_{2k_{FP}}=2i \eta_z \eta_y e^{i(\frac{2}{\sqrt{3}}\varphi_\rho-\frac{\varphi_{2}}{\sqrt{6}}-\frac{\varphi_{1}}{\sqrt{2}})} e^{i P (\frac{2}{\sqrt{3}}\theta_\rho-\frac{\theta_{2}}{\sqrt{6}}-\frac{\theta_{1}}{\sqrt{2}})},~~~~~~~\\
&& S^{y,a}_{2k_{FP}}=2i \eta_x \eta_z e^{i(\frac{2}{\sqrt{3}}\varphi_\rho-\frac{\varphi_{2}}{\sqrt{6}}+\frac{\varphi_{1}}{\sqrt{2}})} e^{i P (\frac{2}{\sqrt{3}}\theta_\rho-\frac{\theta_{2}}{\sqrt{6}}+\frac{\theta_{1}}{\sqrt{2}})},~~~~~~~\\
&& S^{z,a}_{2k_{FP}}=2i\eta_y \eta_x e^{i  (\frac{2}{\sqrt{3}}\varphi_\rho+\sqrt{\frac{2}{3}}\varphi_{2})}e^{i P (\frac{2}{\sqrt{3}}\theta_\rho+\sqrt{\frac{2}{3}}\theta_2)},~
\end{eqnarray}
where $P=R/L=\pm$. 
\begin{eqnarray}
\Delta [\vec{S}^{a}_{2k_{FP}}] = \frac{1}{3}+\frac{g}{3}+\frac{1}{3g}.~
\end{eqnarray}
We can see that when $g=1$, each scaling dimension is 1, the value in the exactly solvable models with non-interacting partons. In the stable SU(2) MSL, we require $g \leq 1$ and hence
\begin{equation}
\Delta[O_{Q_{-}}] \leq \Delta[O_{Q=0}] \leq \Delta[O_{2k_{FP}}] \leq \Delta[O_{Q_{+}}]
\end{equation}
Besides the observables constructed out of local fermion fields discussed above, there are local physical observables that require non-local expressions in terms of fermion fields similar to the string operator defined in Eq.~(\ref{Orbital:string}).  In this SU(2) case, we can consider the ``rung energy'' operator which is symmetric under leg interchange,
\begin{eqnarray}
\epsilon(X)\equiv \tau^x(X,1) \tau^x(X,4).~
\end{eqnarray}
Considering correlation function of such an operator similar to Eq.~(\ref{rungXX_rungXX}) in the spinless case, we can write schematically in our gauge
\begin{eqnarray}
\nonumber && \tau^x(X,1)\tau^x (X,4)\sim~~~~~~~~~~~\\
 && \sim \prod_{X'<X}  \prod_\alpha (-1) c^\alpha(X',1) c^\alpha (X',4) c^\alpha (X',2) c^\alpha(X',3).~~~~~~~~~
 \end{eqnarray}
Such non-local operator in fermionic language seems very intractable but the expression can be greatly simplified under Bosonization,
 \begin{eqnarray}
\nonumber && \prod_{X'<X}  \prod_\alpha c^\alpha(X',1) c^\alpha (X',4) c^\alpha (X',2) c^\alpha(X',3)~\\
&& \sim e^{\pm i\sum_\alpha [\theta(X) + \pi \bar{n}_\alpha X]}= e^{\pm i[ \sqrt{3} \theta_\rho + 3 k_F X]},~
\end{eqnarray}
where we used the definition of $\theta_\rho$ in Eq.~(\ref{Def: SU(2) bosons}), $k_{FR}\equiv k_{F}$ and $\bar{n}_\alpha=k_{F}/\pi$ is the average density of $\alpha$-species fermion. Thus, we can write a contribution to the leg-symmetric energy observable as
\begin{eqnarray}
\epsilon_{3k_{FR}} \sim e^{i \sqrt{3} \theta_{\rho}},~
\end{eqnarray}
with scaling dimension $\Delta[\epsilon_{3k_{FR}}]=\frac{3g}{4}$ and $\epsilon_{3k_{FL}}=\epsilon_{3k_{FR}}^{\dagger}$.  We can also consider other rung energy operator such as $\tau^y(X,1) \tau^y(X,4)$, but the long-wavelength description of such an operator is qualitatively the same as the above $\tau^x(X,1)\tau^x (X,4)$.  
Finally, these local energy observables can be combined with any observables listed earlier to produce further critical operators with potentially enhanced scaling dimension, e.g.\ $O^s_{k_{FR} + 2k_{FL}} \sim \epsilon_{3k_{FL}} O^s_{Q-}$ with $\Delta[O^s_{k_{FR}+2k_{FL}}]=\frac{2}{3}+\frac{g}{12}$ and $\vec{S}^a_{2k_{FR} + 3k_{FL}} \sim \epsilon_{3k_{FL}} \vec{S}^a_{2k_{FR}}$ with $\Delta[\vec{S}^a_{2k_{FR} + 3k_{FL}}] = \frac{1}{3}+\frac{g}{12}+\frac{1}{3g}$.

\section{Zeeman magnetic field effects on the SU(2) Majorana spin liquid}\label{Appendix:Zeeman}
In the SU(2) MSL phase, Zeeman magnetic field only couples to $f^{x}$ and $f^{y}$ fermions, and we can define $f^{\dagger}_{\pm}=(f^{x\dagger}\pm i f^{y\dagger})/\sqrt{2}$ which carry $S^z = \pm 1$ and get Zeeman-shifted, while $f^{z\dagger}$ carries $S^z=0$ that remains unaltered. The spin SU(2) rotation symmetry is broken and only $S^z$ is conserved.  Using symmetry arguments, we can write general four-fermion perturbations in terms of long-wavelength right-moving and left-moving complex fermions as
\begin{eqnarray}
\mathcal{H}_{int}&=&\frac{1}{2} \sum_{\mu,\nu}\lambda^{\mu \nu} \left( \rho_{\mu,R}\rho_{\nu, L} + \rho_{\mu,L}\rho_{\nu,R}\right) \\
&& + w^{+-}(f_{+,R}f_{+,L}f_{-,R}f_{-,L}+\Hc) ~,
\end{eqnarray}
with $\rho_{\mu,P} \equiv f^{\mu\dagger}_P f^\mu_P$, $\mu= +,~-,~z$, and $P=R/L$. The differential RG equations are
\begin{eqnarray}
&& \dot{\lambda}^{++}= - \frac{(w^{+-})^2}{2\pi v_-},~\label{w+-}\\
&& \dot{\lambda}^{--}= - \frac{(w^{+-})^2}{2\pi v_{+}},~\\
&& \dot{\lambda}^{+-} = - \frac{(w^{+-})^2}{\pi(v_+ + v_-)},~\\
&& \dot{w}^{+-} = -\frac{w^{+-}}{2\pi}\left[ \frac{\lambda^{++}}{v_+} + \frac{\lambda^{--}}{v_-} + \frac{4 \lambda^{+-}}{v_+ + v_-}\right],~\\
&& \dot{\lambda}^{zz}=\dot{\lambda}^{+z}=\dot{\lambda}^{-z}=0.~\label{lambdaz}
\end{eqnarray}
Here $\dot{O}\equiv dO/d\ell$, where $\ell$ is logarithm of the length scale and $v_{\pm}$ represent Fermi velocities of the $f^{\pm}$ bands. We see that the MSL is stable if  
\begin{eqnarray}
\frac{\lambda^{++}}{v_{+}} + \frac{\lambda^{--}}{v_{-}} +  \frac{4\lambda^{+-}}{v_{+} + v_{-}}  >0.
\end{eqnarray}
Comparing the RG equations (\ref{w+-})-(\ref{lambdaz}) in the presence of the Zeeman magnetic field with those Eqs.~(\ref{urho})-(\ref{w4}) without the Zeeman field, we see that the instabilities in the ``spin'' sector, $u_{\sigma1}$ and $u_{\sigma2}$, are removed by the magnetic field, and the couplings that contain both $f^\pm$ and $f^z$ do not flow (the reason is that interactions that could cause these to flow do not conserve $S^z$ and thus are not allowed).  An interesting fact about these RG equations is that the instabilities only occur in the $f^{\pm}$ fermion but not in the $f^z$ channel.  Hence, the gapless $f^z$ partons are always gapless no matter how large the Zeeman magnetic field is and always give metal-like contribution to specific heat and thermal conductivity.

\section{SU(2) Majorana spin liquid with Time Reversal Breaking (TRB)}\label{Appendix:TRB}

\begin{figure}[t]
\subfigure[]{\label{2legladder_TRB}\includegraphics[scale=0.5]{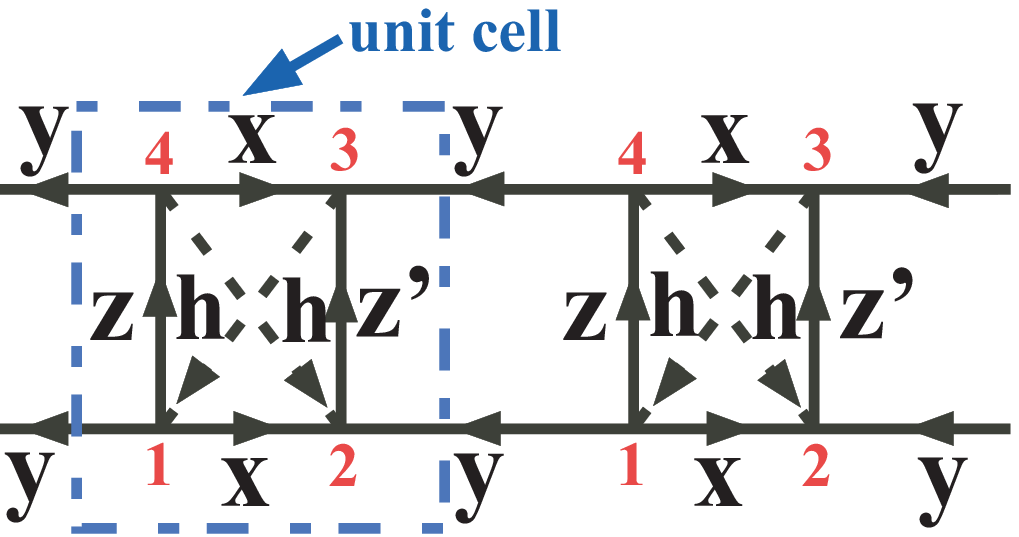}}
\subfigure[]{\label{2legladder_TRB_spec}\includegraphics[scale=0.4]{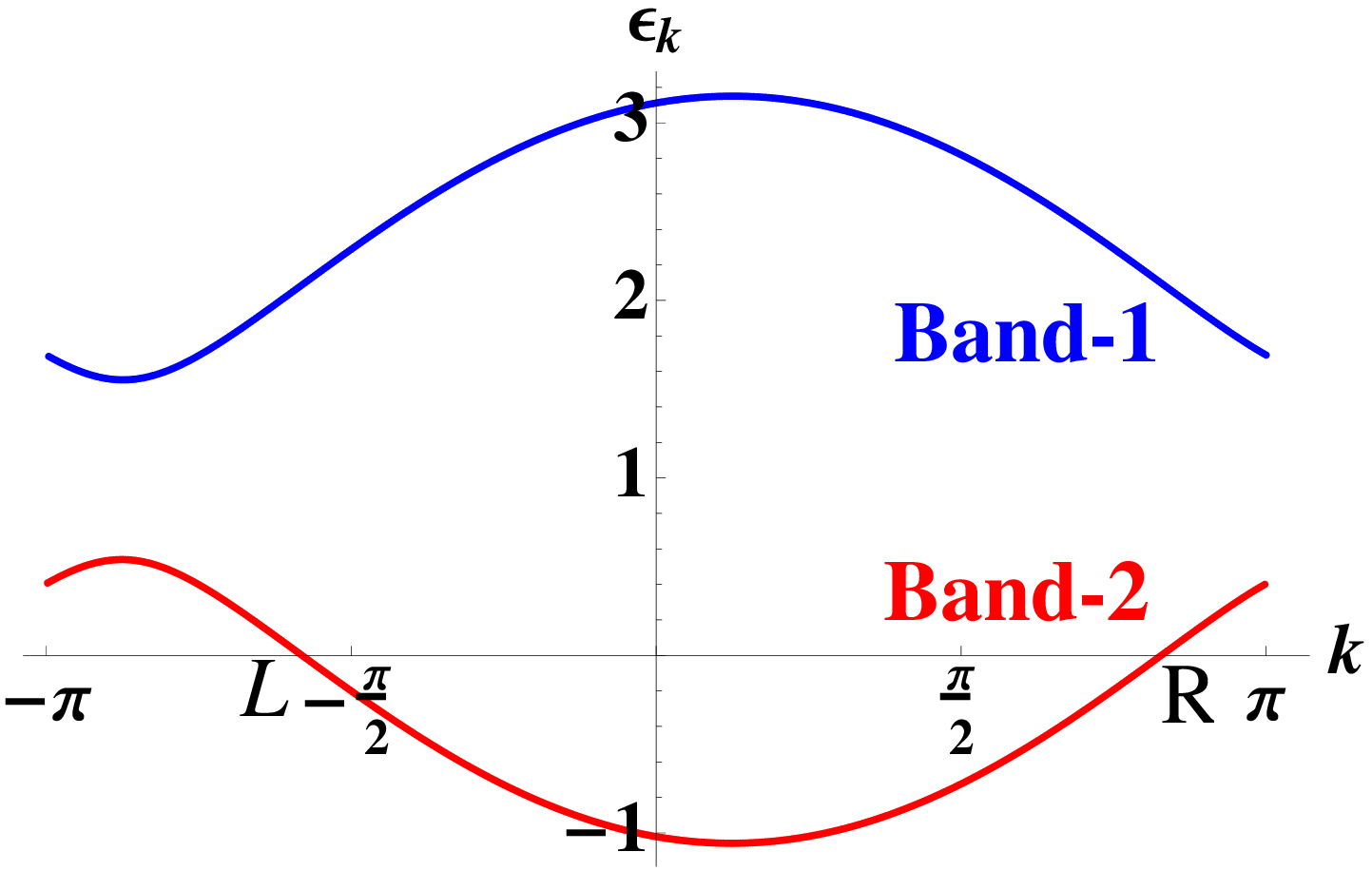}}
\caption{(a) Graphical representation of the exactly solvable Kitaev-type model with time reversal breaking (TRB) introduced by hand and its solution in the zero flux sector. (b) Complex fermion spectrum, Eq.~(\ref{usual_fermion_H}), for the Majorana spin liquid with TRB with $\{J_x,~J_y,~J_z,~J'_z,~h\}=\{1.2,~0.8,~1.0,~1.1,~0.5\}$.}
\label{fig:TRBreaking}
\end{figure}

In this Appendix, we will break the time reversal symmetry explicitly by including a term,
\begin{eqnarray}
&& \mathcal{H}_{TRB} = \frac{h}{2}\sum_{\square_{xz}} \bigg{[} \left(\tau^{x}_1 \tau^{y}_2 \tau^{z}_3 - \tau^{x}_3 \tau^{y}_4 \tau^z_1\right) \left( \vec{\sigma}_3 \cdot \vec{\sigma}_1\right)~\\
&& \hspace{2.5cm}+\left(  \tau^z_2\tau^y_3\tau^x_4 -\tau^z_4 \tau^y_1 \tau^x_2 \right)\left( \vec{\sigma}_{4}\cdot\vec{\sigma}_{2}\right)\bigg{]}.~\label{HTRB}
\end{eqnarray}
Later we will see that such terms reduce the number of four-fermion interactions due to momentum conservation.  Using the Majorana representation, this term can be rephrased as
\begin{eqnarray}
&& \mathcal{H}_{TRB} = i \frac{h}{2} \sum_{\square_{xz}} \bigg{[} \left(\hat{u}_{34} \hat{u}_{41} + \hat{u}_{12} \hat{u}_{23} \right) \sum_{\alpha=x,y,z} c^\alpha_3 c^\alpha_1 \\
&& \hspace{2.4cm} - \left(\hat{u}_{41} \hat{u}_{12} + \hat{u}_{23} \hat{u}_{34} \right) \sum_{\alpha=x,y,z} c^\alpha_4 c^\alpha_2 \bigg{]}.~~~~~
\end{eqnarray}
The graphical representation is shown in Fig.~\ref{2legladder_TRB}.  Before we proceed, we remark that in this case with TRB, we do not need any symmetry to protect the gaplessness, unlike the time reversal invariant case.  The bilinear term $I_{RL}$ that could open a gap is not allowed in the Hamiltonian due to momentum conservation, see below.  For illustration and simplicity, we proceed to take the same parameters as in Sec.~\ref{Sec:SU(2)_case} and include $h$, $\{ J_x,~J_y,~J_z,~J'_z,~h\} = \{ 1.2,~0.8,~1.0,~1.1,~0.5 \}$. The complex fermion spectrum is shown in Fig.~\ref{2legladder_TRB_spec}, and we can clearly see that due to the presence of the Time-Reversal Breaking term, there is no Right-Left symmetry anymore (i.e.\ $k_{FL} \neq -k_{FR}$).  In the weak-coupling regime, the general four-fermion interactions can be written as
 \begin{eqnarray}
\mathcal{H}_{int}^{TRB}=\tilde{u}_{\rho}\mathcal{J}_R \mathcal{J}_L - \tilde{u}_{\sigma 1} \vec{\mathcal{J}}_R \cdot \vec{\mathcal{J}}_L + \tilde{u}_{\sigma 2} I^{\dagger}_{RL} I_{RL},~
\end{eqnarray}
where $\mathcal{J}_P$, $\vec{\mathcal{J}}_P$, and $I_{RL}$ are defined in Eqs.~(\ref{Jrho})-(\ref{Irl}). We can see that the number of allowed interactions is reduced because there is no special relation between $k_{FR}$ and $k_{FL}$ and additional terms are forbidden by momentum conservation.

The weak-coupling differential RG equations in this case are
\begin{eqnarray}
&& \dot{\tilde{u}}_{\rho} = \frac{1}{\pi (v_R+v_L)}\left[ \tilde{u}_{\sigma 2}^2+2 \tilde{u}_{\sigma 1} \tilde{u}_{\sigma 2} \right],~\\
&& \dot{\tilde{u}}_{\sigma 1} = \frac{1}{\pi (v_R+v_L)} \left[ -\tilde{u}_{\sigma 1}^2 + 2 \tilde{u}_{\sigma 1} \tilde{u}_{\sigma 2} \right],~\\
&& \dot{\tilde{u}}_{\sigma 2} = \frac{1}{\pi (v_R+v_L)} \left[ -3\tilde{u}_{\sigma 2}^2-6 \tilde{u}_{\sigma 1} \tilde{u}_{\sigma 2}\right].~
\end{eqnarray}
We can give a qualitative description of the stable flows.\cite{Itoi97}  If $\tilde{u}_{\sigma 1} > 0$ and $\tilde{u}_{\sigma 1} + \tilde{u}_{\sigma2} > 0$, the couplings $\tilde{u}_{\sigma 1,2}$ are marginally irrelevant and flow to zero, $u_{\sigma1}^*=u_{\sigma2}^*=0$.  The coupling $\tilde{u}_\rho$ approaches a fixed value, $\tilde{u}^*_\rho$, and is strictly marginal; 
unlike the time reversal symmetric case in Sec.~\ref{Sec:SU(2)_case}, there is no condition on the sign of $\tilde{u}^*_\rho$.
We conclude that the SU(2) MSL with explicit time reversal breaking is stable in a wide regime of parameters. We also note that, even though initially there is no conservation of the $f$-fermions in this model, breaking TRS leads to $k_{FL} \neq -k_{FR}$ and prohibits four-fermion interactions such as $f^\alpha f^\beta f^\gamma f^\delta$ and $f^{\alpha\dagger} f^\beta f^\gamma f^\delta$, so the fermion conservation emerges at low energy.  We note that if we rewrite the couplings as
\begin{eqnarray}
&& \tilde{u}_{\sigma 1}= - \frac{\pi(v_R+v_L)}{2\sqrt{2}}g_1,~\\
&& \tilde{u}_{\sigma 2}=\frac{\pi(v_R+v_L)}{2\sqrt{2}} ( g_1 +g_2),~
\end{eqnarray}
the RG equations can be rephrased as
\begin{eqnarray}
&& \dot{\tilde{g}}_{\rho}=3 \dot{\tilde{u}}_{\rho}+\dot{\tilde{u}}_{\sigma2}=0,~\\
&& \dot{g}_{1}=\frac{1}{2\sqrt{2}} \left( 3 g^2_{1}+  2g_1g_2 \right),~\\
&& \dot{g}_{2}=\frac{1}{2\sqrt{2}} \left( -3 g^2_{2}- 2 g_1 g_2 \right).~
\end{eqnarray}
The last two equation are exactly the same as one-loop RG equations in an SU(3) WZW model in Ref.~\onlinecite{Itoi97}.  Note that in the SU(2) MSL the ``charge'' ($\rho$) sector also remains gapless, cf.\ Appendix~\ref{Appendix:SU(2)observables}.
\acknowledgments
This research is supported by the National Science Foundation through grant DMR-0907145 and by the A.~P.~Sloan Foundation.

\bibliography{biblio4MajoranaSL}

\begin{thebibliography}{57}
\bibitem [{BC()}]{BC}%
  \BibitemOpen
  \href@noop {} {}\bibinfo {note} {In the calculations, we used anti-periodic
  boundary conditions in the horizontal direction, which gives exact solution
  in the sector with $\prod_{j \in leg-1}\tau^{z}_{j}=\prod_{j \in
  leg-2}\tau^{z}_{j}=1$}\BibitemShut {NoStop}%
\end{thebibliography}
\end{document}